\documentclass[aps,prd,reprint,superscriptaddress,nofootinbib]{revtex4-2}

\usepackage{graphicx}
\usepackage{subcaption}
\usepackage{amssymb,amsmath, amsfonts}
\usepackage{adjustbox}
\usepackage{lipsum}
\usepackage{epsfig}
\usepackage{xcolor}
\usepackage{tabularx}
\usepackage{soul}
\usepackage{multirow}
\usepackage{lineno,hyperref}

\usepackage{inputenc}
\usepackage{booktabs}
\usepackage{float}

\usepackage{lmodern}

\newcommand{\apjl}{Astrophys.\ J.\ Lett.}

\newcommand{\mnras}{Mon.\ Not.\ R.\ Astron.\ Soc.}

\newcommand{\aap}{Astron.\ Astrophys.}

\newcommand{\apss}{Astrophys.\ Space Sci.}

\usepackage{microtype}
\begin{document}

\title{Indication for Dual Periodic Signatures in PKS\,0805-07 from Multi-technique Time Series Analysis }

\author{Sikandar Akbar}
\email{darprince46@gmail.com}
\affiliation{Department of Physics, University of Kashmir,  Srinagar-190006, India}

\author{Zahir Shah}
\email{shahzahir4@gmail.com}
\affiliation{Department of Physics, Central University of Kashmir, Ganderbal-191131, India}

\author{Ranjeev Misra}
\affiliation{Inter-University Centre for Astronomy and Astrophysics, Post Bag 4, Ganeshkhind, Pune-411007, India}

\author{Sajad Boked}
\affiliation{Department of Physics, University of Kashmir,  Srinagar-190006, India}

\author{Naseer Iqbal}
\affiliation{Department of Physics, University of Kashmir,  Srinagar-190006, India}

\begin{abstract}
We report the identification of two statistically significant quasi-periodic oscillations in the weekly binned $\gamma$-ray light curve of the flat-spectrum radio quasar PKS\,0805-07, observed by Fermi-LAT over the period MJD 59047.5-59740.5. By applying a suite of complementary time-series analysis techniques, we identify periodic signatures at $\sim$255 and $\sim$112 days. These techniques include the Lomb-Scargle periodogram (LSP), Weighted Wavelet Z-transform (WWZ), REDFIT, Date-Compensated Discrete Fourier Transform (DCDFT), Phase Dispersion Minimization (PDM), and the String-length method. The reliability of these signals is supported by high local significance ($\geq 99\%$) in all methods and reinforced through phase-folding. Model selection using the Akaike Information Criterion (AIC) and Bayesian Information Criterion (BIC) strongly supports a two-component periodic model.
The detection of dual QPOs is rare among blazars and suggests complex variability mechanisms. Although a binary supermassive black hole (SMBH) scenario could be considered given the source’s high redshift ($z = 1.837$), the short periodicities are difficult to reconcile with orbital motion unless invoking extreme parameters.
Double-sine model fitting reveals that the oscillatory components have comparable amplitudes but are out of phase, suggesting a potential beating phenomenon due to interference. This amplitude-modulated variability is consistent with a
geometric origin, most plausibly jet precession driven by Lense-Thirring torques, superimposed with
a secondary process such as polar jet oscillation. Doppler factor modulation arising from these effects
can account for the observed flux variations without requiring an unrealistically compact binary.
\end{abstract}

\keywords{galaxies: active - galaxies:  BL Lacertae objects: PKS\,0805-07 - galaxies: jets - radiation mechanisms: non-thermal - gamma-rays: galaxies.}
\maketitle

\section{Introduction}
\label{introduction}
PKS\,0805-07, a high-redshift flat-spectrum radio quasar (FSRQ), has drawn considerable attention in recent years due to its pronounced $\gamma$-ray activity and recurrent flaring episodes. Flat Spectrum Radio Quasars (FSRQs) belong to the broader class of blazars-active galactic nuclei (AGN) characterized by relativistic jets pointed close to our line of sight. This orientation makes them ideal laboratories for studying extreme variability and jet-related phenomena. In our previous work \citep{2024ApJ...977..111A}, we performed a comprehensive temporal and spectral analysis of PKS\,0805-07 using Fermi-LAT and Swift observations spanning over 14 years. Building on this foundation, we now investigate quasi-periodic oscillations (QPOs) in its $\gamma$-ray light curve. The search for QPOs in blazars provides a powerful diagnostic for probing processes near supermassive black holes (SMBHs), including jet precession, disk instabilities, and binary black hole interactions.

In a systematic all-sky search using Fermi-LAT data, \citet{2017MNRAS.471.3036P} reported evidence for long-term periodic behavior in the $\gamma$-ray light curves of four blazars, among which PKS\,0805-07 emerged as a noteworthy candidate. This FSRQ, located at a redshift of $z=1.837$, was found to exhibit a  periodicity of approximately 658 days with a high significance under white-noise assumptions ($\sim 5\sigma$). Although the significance under red-noise modeling was relatively lower ($\sim 93.3\%$), the detection was further supported by independent likelihood-based analyses, which identified a similar periodicity at $\sim$676 days. These findings raised the possibility that PKS\,0805-07 could host a binary SMBH system or exhibit jet-related precession, aligning with theoretical scenarios proposed for quasi-periodic blazar behavior.

However, establishing the authenticity of QPOs in AGN light curves demands a rigorous treatment of the underlying stochastic processes that give rise to red-noise-dominated variability. Simple white-noise assumptions can often overestimate the significance of periodic signals. To overcome these limitations, more advanced methodologies involving stochastic modeling and simulation-based inference have gained prominence. In particular, first-order autoregressive (AR(1)) modeling have proven effective in distinguishing true periodic components from spurious fluctuations arising from correlated noise.

\begin{figure*}
    \centering
    \includegraphics[scale=0.4,angle=0]{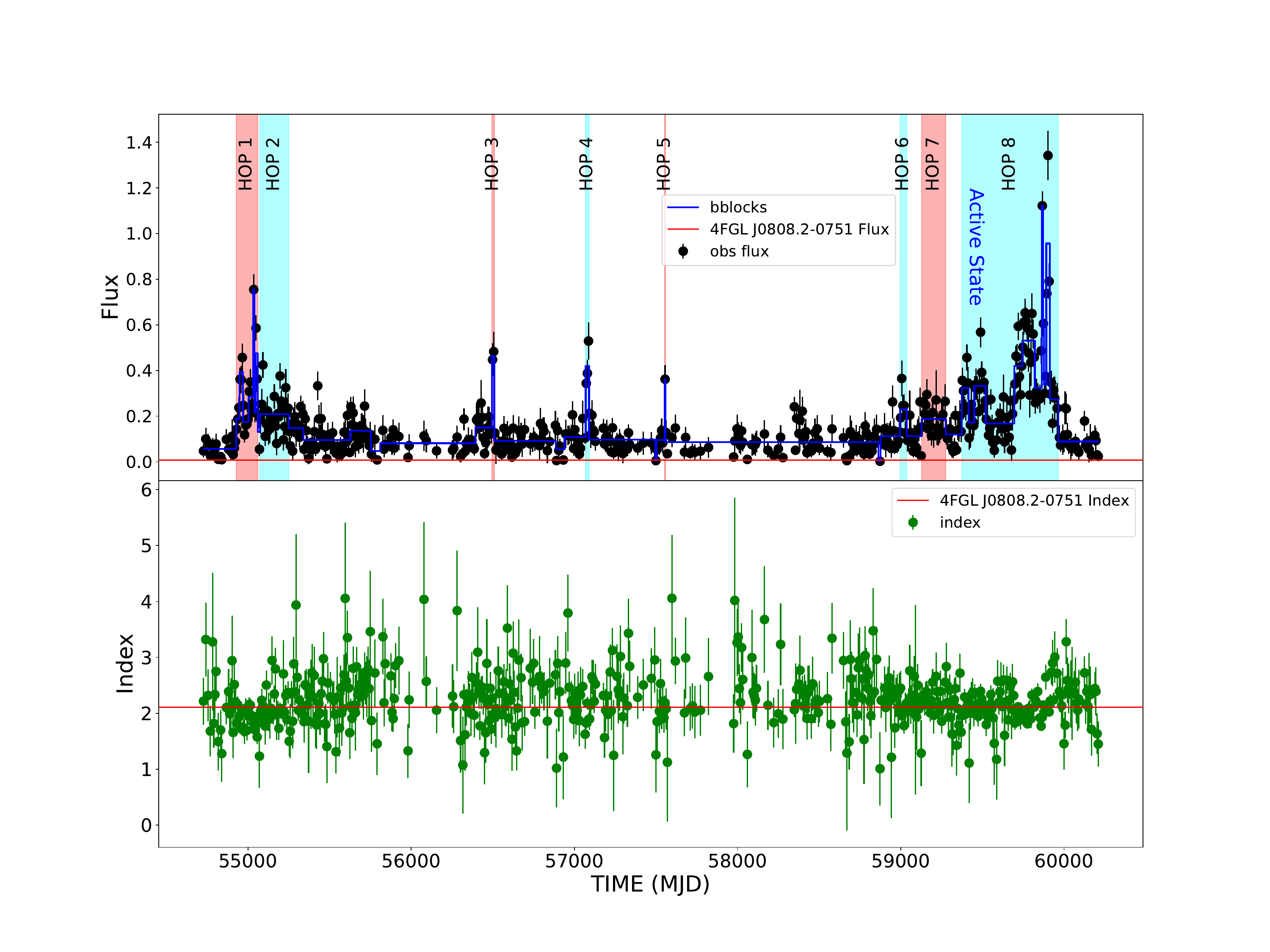}
    \caption{The upper panel depicts the weekly binned $\gamma$-ray light curve of PKS\,0805-07, integrated over the energy range 0.1-100 GeV [Flux ($E > 100$ MeV)] in units of $10^{-6}\,\text{photons\,cm}^{-2}\,\text{s}^{-1}$ from MJD 54684 to 60264. The lower panel shows the corresponding spectral index values for the same period. Red horizontal lines in both panels represent the flux and index reported in the 4FGL catalog. Shaded regions denote the HOP intervals, with HOP 8 designated as the ``active state''. Different HOPs are color-coded for clarity. Reproduced from  \citet{2024ApJ...977..111A}. The time interval MJD 59047.5-59740.5, used for the present QPO study, spans the rise and peak of HOP 8, including the pre-flare phase.}
    \label{bb_lc}
\end{figure*}

In this work, we revisit the $\gamma$-ray variability of PKS\,0805-07 using high-cadence Fermi-LAT light curve to perform an in-depth periodicity analysis. In our earlier study \citep{2024ApJ...977..111A}, we followed the approach outlined in \citep{2019ApJ...877...39M}, which combines the Bayesian Blocks (BB) algorithm \citep{2013ApJ...764..167S} with the HOP  algorithm \citep{1998ApJ...498..137E}. The BB algorithm partitions the light curve into segments of statistically significant flux variation, while the HOP algorithm, originally developed for identifying particle groupings in cosmological simulations, assigns each data point to its densest neighboring point via an iterative hill-climbing procedure that follows the gradient to local density maxima. When adapted to time-domain analysis, HOP clusters adjacent high-flux BB segments into coherent structures termed “HOP groups,” each representing a physically meaningful activity phase in the light curve. Using this framework, we identified HOP~8 (MJD~59370–59965) as the dominant flaring episode (active state), characterized by significantly elevated flux levels compared to prior intervals (see Figure \ref{bb_lc}).

In the present study, we focus our analysis on the interval MJD~59047.5–59740.5. This time range was selected because it encompasses the portion of the light curve where the double-periodic quasi-periodic oscillation like behavior is most prominent, and the variability pattern is least affected by data gaps or prolonged quiescence. The window includes multiple complete cycles  periodicities, thereby allowing a statistically meaningful assessment of the periodic modulations. Furthermore, this interval captures both the rise and peak phases of HOP~8, as well as the immediate pre-flaring activity, ensuring that we trace the evolution of variability leading into the dominant outburst. This selection provides a statistically optimal baseline for probing QPO signatures linked to jet dynamics or underlying emission mechanisms.\\

We apply multiple time-series techniques-including Lomb-Scargle periodograms (LSP), weighted wavelet Z-transforms (WWZ), phase--folding analysis,  Phase Dispersion Minimization (PDM) and string-length methods to assess the consistency and physical plausibility of the claimed QPO. Our goal is to place the periodicity claims on a statistically sound footing while accounting for red noise, uneven sampling, and finite-duration effects. The results not only test the periodicity hypothesis for PKS\,0805-07 but also provide a framework for future investigations of similar blazar candidates.

The structure of this paper is as follows: Section~\ref{sec:1} describes the data selection and reduction. Section~\ref{QPO} presents the results of various time-series analyses applied to the $\gamma$-ray light curve. Section~\ref{evo} examines the temporal evolution of the detected QPOs, while Section~\ref{sine_fit} details the double-sine model fitting. Section~\ref{beat}  discusses the amplitude-modulated jet precession model. Finally, Section~\ref{sum} summarizes the findings and discusses their physical implications.

\section{Observations and Data reduction}
\label{sec:1}

\subsection{Fermi-LAT}
The Fermi Large Area Telescope (Fermi-LAT), part of the Fermi Gamma-ray Space Telescope mission (formerly GLAST), is a high-energy space-based observatory launched by NASA in 2008. It features a broad field of view of about 2.3 steradians and typically operates in scanning mode, enabling full-sky coverage in the energy range of approximately 20\,MeV to 500\,GeV every three hours \citep{2009ApJ...697.1071A}.
Data reduction and analysis were performed using \texttt{Fermitools} version 2.2.0, available through the Anaconda Cloud via the Fermi Science Support Center (FSSC). We followed the standard data processing protocols as outlined in the official Fermi-LAT documentation\footnote{\url{https://fermi.gsfc.nasa.gov/ssc/data/analysis/}}. Photon events from the SOURCE class (\texttt{evclass=128}, \texttt{evtype=3}) were selected within a 15\textdegree{} region of interest (ROI) centered on the target source, applying a zenith angle cut of 90\textdegree{} to minimize contamination from Earth's limb emission.
For spectral analysis, we used photons in the 0.1--300\,GeV energy range and performed additional analysis with \texttt{FERMIPY} (v1.0.1) \citep{2017ICRC...35..824W}\footnote{\url{https://fermipy.readthedocs.io/en/latest/}}. The Galactic diffuse background was modeled using \texttt{gll\_iem\_v07.fits}, while the isotropic component was represented by \texttt{iso\_P8R3\_SOURCE\_V3\_v1.txt}. We adopted the post-launch instrument response function \texttt{P8R3\_SOURCE\_V3}.
The model included all 4FGL catalog sources within a 25\textdegree{} ROI centered on the source. For sources within 10\textdegree{}, normalization parameters were left free during the fit. For PKS\,0805$-$07, both the normalization and spectral shape parameters ($\alpha$ and $\beta$) were allowed to vary, while the remaining sources had their parameters fixed to the catalog values. \par

We adopted criteria to filter out the sources with low Test Statistics (TS) i.e. below TS = 9 and generated weekly binned $\gamma$-ray light curve of the source of interest with TS$\> (\ge 9)$. In this study, we analyze the time interval MJD 59047.5-59740.5, which fully covers the rise and main phase of the HOP 8 activity, as well as the preceding pre-flare stage.

\section{Quasi-periodic oscillation}
\label{QPO}
To investigate the presence of QPOs in the $\gamma$-ray light curve of the blazar PKS\,0805-07, we employed a suite of time-series analysis techniques. These include the LSP, WWZ, REDFIT based on a first-order autoregressive process [AR(1)], DCDFT, PDM method, and the string-length method. In addition, we applied a double-sine model fit and performed a Monte Carlo simulation test to assess the statistical significance of the detected signals. A detailed description of each method, along with the corresponding results, is presented in the following sections.

\begin{figure*}
    \centering
    \includegraphics[width=0.9\textwidth]{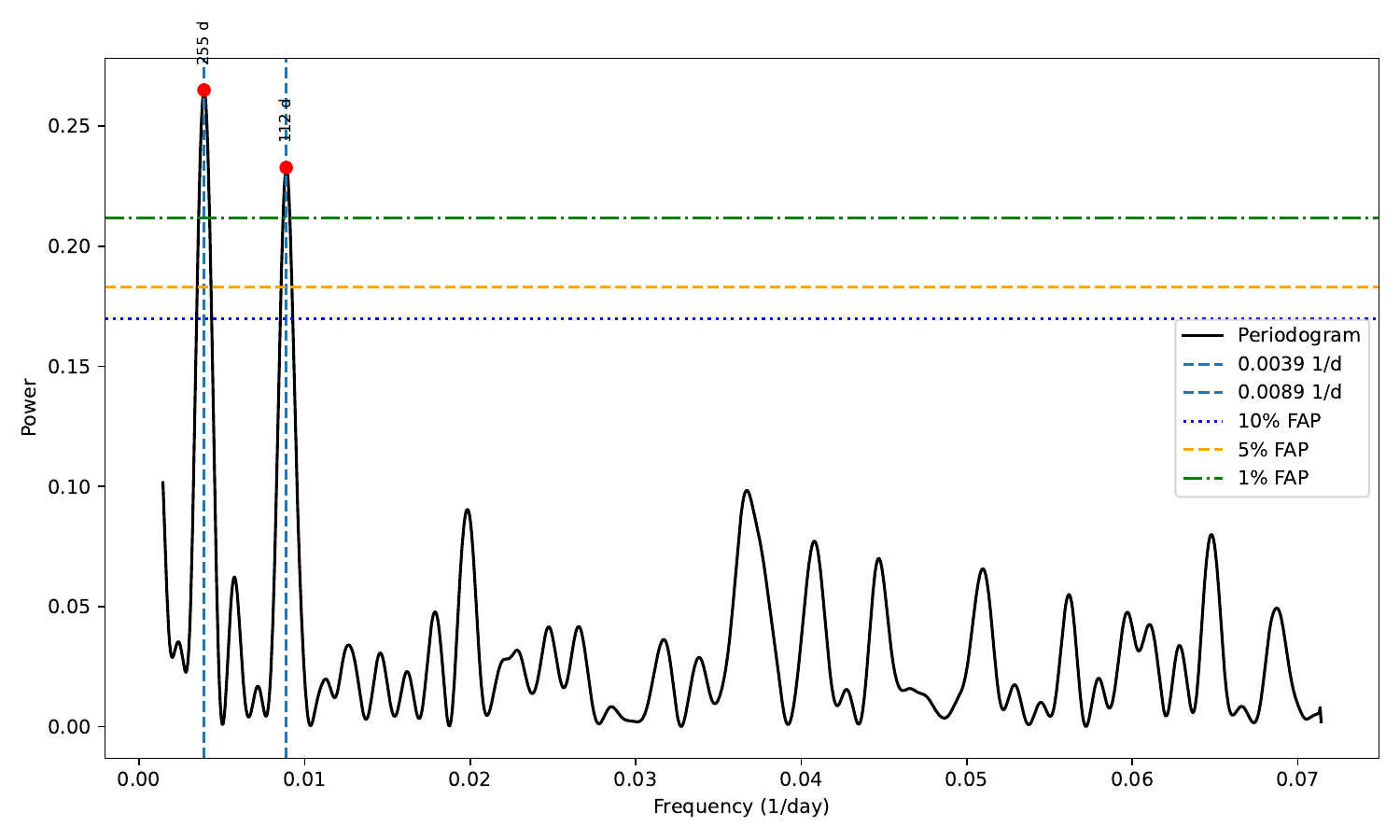}
    \caption{LSP of the $\gamma$-ray light curve of PKS\,0805-07. The two most prominent peaks are observed at frequencies of $0.0039$ and $0.0089$~day$^{-1}$, corresponding to periods of approximately $255$ and $112$ days, respectively. The horizontal lines represent the false alarm probability (FAP) thresholds at 10\% (blue dotted), 5\% (orange dashed), and 1\% (green dash-dotted), indicating the statistical significance of the detected periodicities.}

    \label{LSP}    
\end{figure*}

\subsection{Lomb-Scargle Periodogram (LSP)}\label{sec:lsp}
The LSP is a widely adopted technique for detecting periodic signals in time-series data, particularly when the observations are unevenly sampled \citep{lomb1976least, scargle1982studies}. This method is especially effective for analyzing non-uniformly spaced light curves to uncover underlying periodicities. In this study, we employed the \textsc{LOMB-SCARGLE}\footnote{\url{https://docs.astropy.org/en/stable/timeseries/lombscargle.html}} implementation from the \textsc{ASTROPY} library to carry out the analysis.  The uncertainties in the flux measurements were incorporated to enhance the reliability of the resulting periodogram.

The LSP computes the normalized spectral power $P_{LS}(f)$ at a given frequency $f$ using the expression \citep{vanderplas2018understanding}:

\begin{equation}
\begin{split}
P_{LS}(f) = \frac{1}{2} \bigg[ &\frac{\left(\sum_{i=1}^{N} g_i \cos(2\pi f (t_i - \tau))\right)^2}{\sum_{i=1}^{N} \cos^2(2\pi f (t_i - \tau))} \\
&+ \frac{\left(\sum_{i=1}^N g_i \sin(2\pi f (t_i - \tau))\right)^2}{\sum_{i=1}^N \sin^2(2\pi f (t_i - \tau))} \bigg]
\end{split}
\end{equation}

where the phase offset $\tau$ is given by:

\begin{equation}
    \tau = \frac{1}{4\pi f}\tan^{-1} \left( \frac{\sum_{i=1}^N \sin(4\pi f t_i)}{\sum_{i=1}^N \cos(4\pi f t_i)} \right)
\end{equation}
In our analysis, the minimum and maximum search frequencies ($f_{\rm min}$ and $f_{\rm max}$) were selected as $1/T$ and $1/(2\Delta T)$, respectively, where $T$ is the total duration of the observations and $\Delta T$ is the bin size or average time between consecutive data points. These limits define the frequency range investigated by the LSP.

The analysis revealed two statistically significant peaks in the periodogram at frequencies of $0.003924 \pm 0.000409\ \text{day}^{-1}$ and $0.008920 \pm 0.000463\ \text{day}^{-1}$, corresponding to periods of approximately $254.8 \pm 26.5$ days and $112.1 \pm 5.8$ days, respectively (see Figure~\ref{LSP}).
The uncertainty in the detected period was estimated by fitting a Gaussian function to the peak of the periodogram and adopting the half-width at half-maximum (HWHM) as the associated error \citep{vanderplas2018understanding, sharma2024detection}. Furthermore, the Generalized Lomb-Scargle periodogram (GLSP) is commonly employed to validate the presence of periodic signatures, as it incorporates measurement uncertainties directly into the analysis. The results obtained from the GLSP further support the presence of periodicity, thereby reinforcing the findings derived from the standard Lomb-Scargle method.

To evaluate the statistical significance of the detected periodicities, We employed the \texttt{LombScargle.false\_alarm\_probability()} implementation from the \texttt{astropy.timeseries} module with \texttt{method="baluev"}, which provides an analytic approximation of the  false alarm probability (FAP). This method is based on the formalism developed by  \citet{2008MNRAS.385.1279B},  which uses extreme value statistics to analytically estimate the FAP while accounting for the trial correction across the entire frequency range scanned. The number of independent frequencies, $N_{ind}$ , is internally estimated based on the time sampling, frequency grid resolution, and baseline of the data, and does not need to be explicitly specified by the user. 
This approach provides a statistically grounded estimate of the probability that a given peak in the periodogram could result purely from stochastic noise fluctuations, thereby enabling a reliable assessment of its significance. The red-noise behavior commonly observed in AGN and blazar light curves is typically attributed to underlying stochastic processes and is well described by a power-law power spectral density (PSD) of the form $P(\nu) \sim A \nu^{-\beta}$, where $\nu$ is the temporal frequency and $\beta > 0$ denotes the spectral index. To assess the statistical significance of the peaks detected in the LSP  analysis, we employed a Monte Carlo simulation technique. Specifically, we generated $1 \times 10^5$ synthetic light curves that replicate both the PSD and the probability density function (PDF) of the observed light curve, following the method proposed by \citet{emmanoulopoulos2013generating}. The local significance of each candidate peak was evaluated by comparing it against the distribution of powers at the corresponding frequencies across the simulated datasets. To address the global significance of these peaks, we computed the global false alarm probability (FAP) for each candidate frequency by comparing its observed LSP power to the distribution of maximum powers obtained across the entire frequency range from 100,000 simulated light curves. This approach accounts for the ``look-elsewhere'' effect. The peak at $0.003924 \ \mathrm{day}^{-1}$ ($\sim255$ days) exhibits a global FAP of 0.020, while the peak at $0.008920 \ \mathrm{day}^{-1}$ ($\sim112$ days) has a global FAP of 0.046. Both peaks thus remain statistically significant at the $\geq 95\%$ global confidence level.
\begin{figure*}
    \centering
    \begin{subfigure}[t]{0.95\textwidth}
        \centering
        \includegraphics[width=\textwidth]{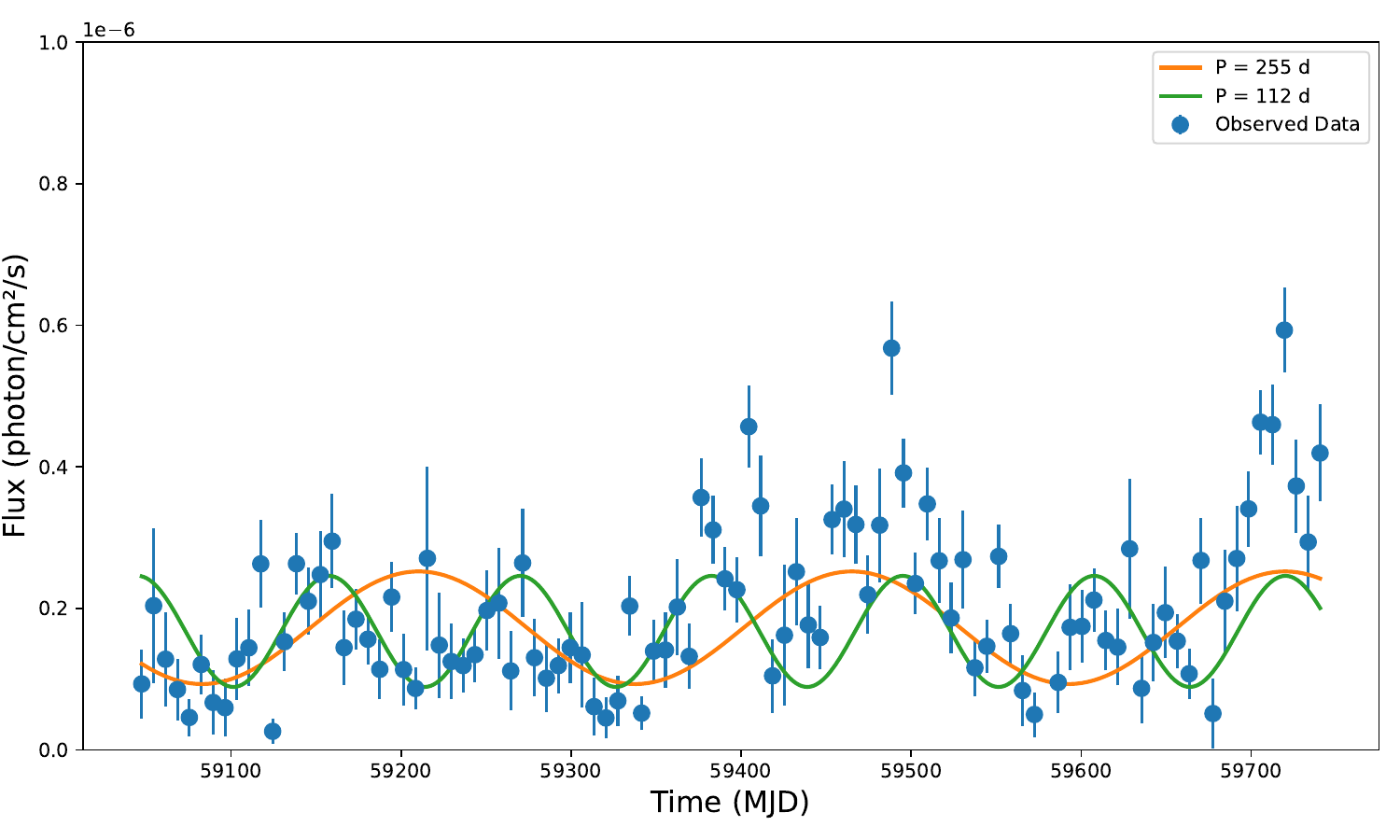}
        \label{sinefit}
    \end{subfigure}
    
    \vspace{0.8em}
    
    \begin{subfigure}[t]{0.95\textwidth}
        \centering
        \includegraphics[width=\textwidth]{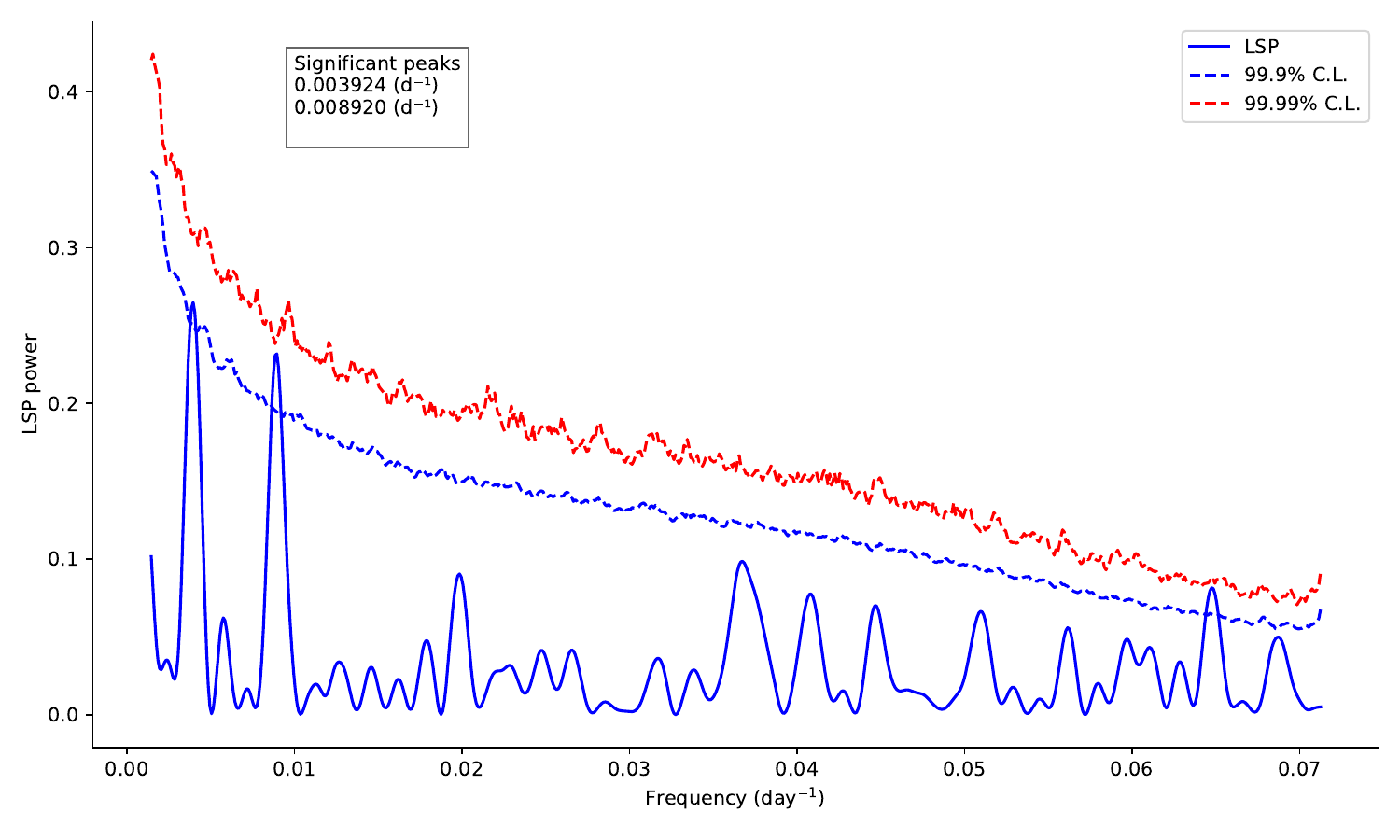}
        \label{LSP_sim}
    \end{subfigure}
    
    \caption{Top: The $\gamma$-ray light curve of PKS\,0805-07 (MJD 59047.5-59740.5), overplotted with the best-fit sinusoidal models corresponding to the two significant periods identified via LSP analysis: \( P_1 = 254.8 \pm 26.5 \) days and \( P_2 = 112.1 \pm 5.8 \) days. Bottom: LSP of the same light curve, showing two dominant peaks at frequencies 0.003924~day$^{-1}$ and 0.008920~day$^{-1}$ (periods $\sim$255 and $\sim$112 days), both exceeding the 99.9\% confidence level derived from $10^5$ Monte Carlo simulations using the method of \citet{emmanoulopoulos2013generating}.}
    \label{fig:sine_lsp_combined}
\end{figure*} \\\\
The LSP analysis revealed two prominent peaks, located at approximately $\sim$0.003924~day$^{-1}$ ($\sim$254.8 days) and $\sim$0.008920~day$^{-1}$ ($\sim$112.1 days), both exhibiting significance levels exceeding 99.9\% (see bottom panel of Figure \ref{fig:sine_lsp_combined}).
 These results strongly support the presence of two quasi-periodic oscillations  in the $\gamma$-ray light curve over the time span MJD~59047.45-59740.5, corresponding to periodicities of $\sim$254.8 days and $\sim$112.1 days. These periods correspond to approximately 2.5 and 6 cycles within the observational window. See the top panel of Figure~\ref{fig:sine_lsp_combined} for a visual representation. To verify that the detected QPOs are not caused by instrumental or environmental effects, we performed a LSP periodogram analysis on the flux uncertainties. We note that the LSP of the flux uncertainties reveals a significant peak at 0.0190~day$^{-1}$ ($\sim$52.6 days), which does not coincide with the QPO frequencies identified in the actual flux data ($\sim$255 and $\sim$112 days). No significant power was observed at or near the QPO timescales, confirming that the periodic signals are intrinsic to the source and not artifacts introduced by the observational setup or satellite operations.

\begin{figure*}
    \centering
    \includegraphics[width=1.0\textwidth]{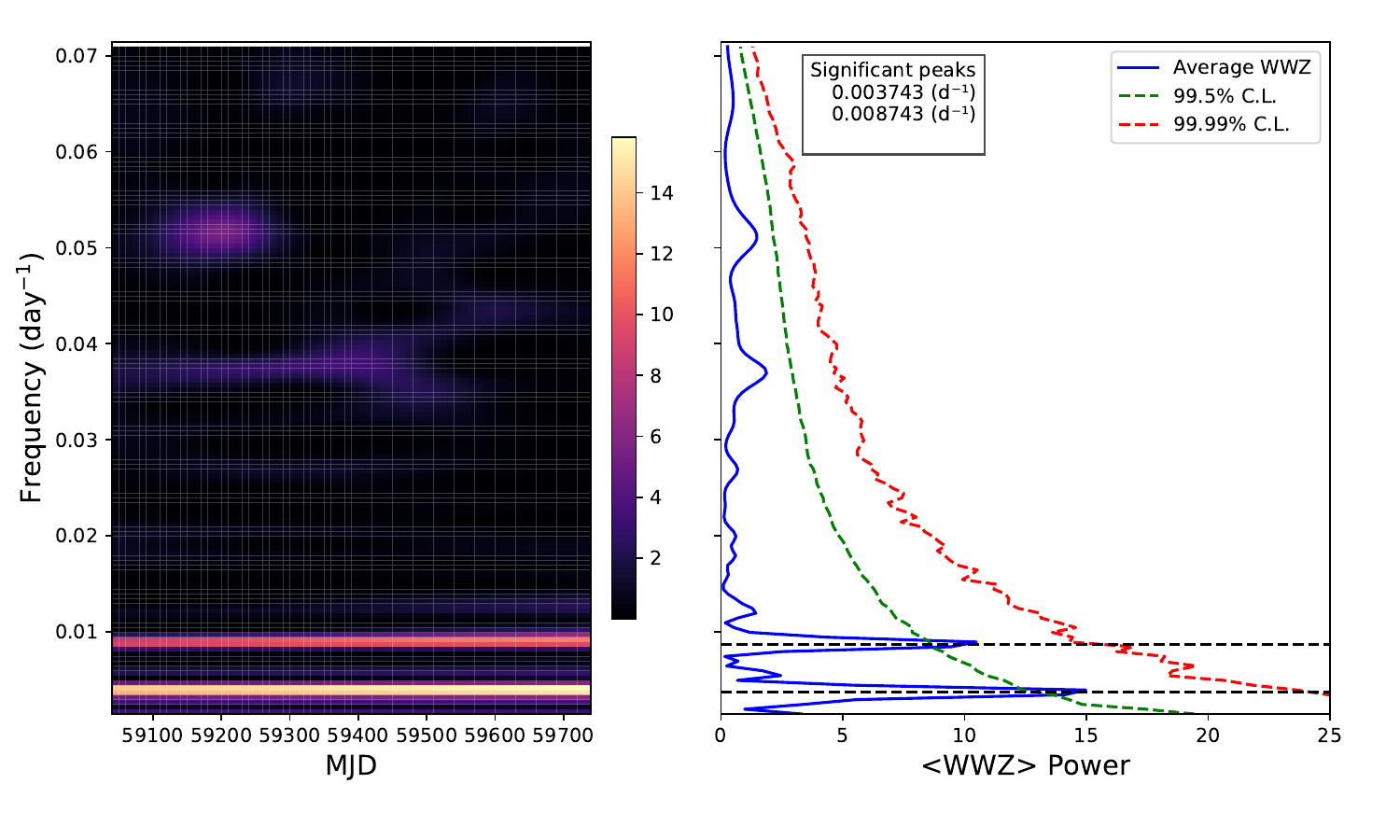}
   \caption{Left: WWZ map of the $\gamma$-ray light curve showing the evolution of power as a function of time (MJD) and frequency. Right: The average WWZ power spectrum with confidence levels derived from $10^5$ Monte Carlo simulations. The dashed green and red lines represent the 99.5\% and 99.99\% confidence levels, respectively. The dotted and dashed black lines mark the two most significant peaks at frequencies of 0.003743 and 0.008743 day$^{-1}$, corresponding to periods of approximately 267 and 114 days.}

    \label{wwz}    
\end{figure*}

\subsection{Weighted Wavelet Z-Transform (WWZ)}\label{sec:wwz}
Wavelet analysis is a powerful approach for identifying periodic components in time-series data by simultaneously resolving signals in both the time and frequency domains. This technique is particularly advantageous for tracking the temporal evolution of QPO features, offering a nuanced view of how periodic signals arise, vary, and diminish over time \citep{foster1996wavelets}.

In our analysis, we adopted the abbreviated Morlet wavelet kernel, mathematically described as:

\begin{equation}
    f[\omega (t - \tau)] = \exp[i \omega (t - \tau) - c \omega^2 (t - \tau)^2]
\end{equation}

The corresponding WWZ projection is expressed as:

\begin{equation}
    W[\omega, \tau: x(t)] = \omega^{1/2} \int x(t)f^* [\omega(t - \tau)] dt
\end{equation}

where $f^*$ denotes the complex conjugate of the wavelet kernel $f$, $\omega$ is the angular frequency, and $\tau$ represents the temporal shift. For implementation, we utilized the publicly available Python-based WWZ\footnote{\url{https://github.com/eaydin/WWZ}} package.

The WWZ power spectrum revealed two notable peaks, one located at a frequency of $0.003743 \pm 0.000596~\mathrm{day^{-1}}$, corresponding to a period of $267.17 \pm 44.89$ days, and the other at $0.008743 \pm 0.000584 ~\mathrm{day^{-1}}$, corresponding to a period of $114.38 \pm 7.64$ days. To evaluate the statistical significance of the peaks identified in the WWZ analysis, we utilized a Monte Carlo simulation approach. A total of $1 \times 10^5$ synthetic light curves were generated, each designed to reproduce both the PSD and PDF of the observed data, following the methodology of \citet{emmanoulopoulos2013generating}. The significance of each candidate peak was determined by comparing its power to the distribution of powers at the corresponding frequency obtained from the simulated light curves. The uncertainties in the period estimates were derived by fitting a Gaussian function to the respective peaks in the average WWZ profile (see Figure \ref{wwz}). 

\begin{figure*}
    \centering
    \includegraphics[width=1.0\textwidth]{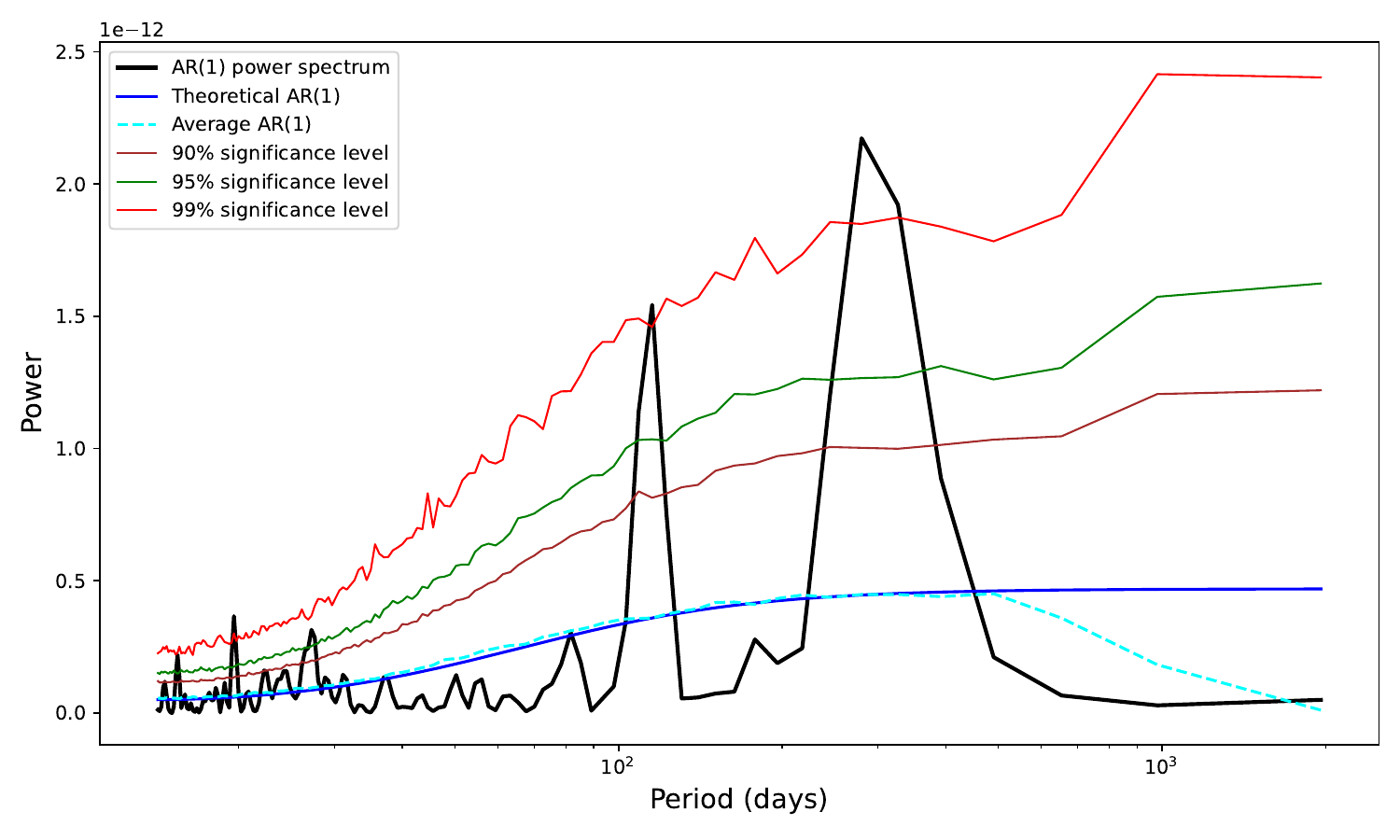}
    \caption{The red noise-corrected REDFIT power spectrum of the $\gamma$-ray light curve of PKS\,0805-07. The black curve shows the AR(1) power spectrum, while the blue and cyan curves represent the theoretical and average AR(1) models, respectively. The brown, green, and red lines denote the 90\%, 95\%, and 99\% Monte Carlo significance levels. Two dominant peaks are observed, both exceeding the 99\% confidence level.}
    \label{Fig-redfit}    
\end{figure*}

\subsection{First-order Autoregressive Process (REDFIT)}\label{sec:redfit}
The variability observed in AGN light curves, including those of blazars, is often governed by red noise-originating from stochastic fluctuations in the jet or accretion disc. Such behavior can be effectively described using a first-order autoregressive process, AR(1), which models the current emission as linearly dependent on its immediate past value \citep{schulz2002redfit}. Mathematically, this relation is expressed as $r(t_i) = A_i r(t_{i-1}) + \epsilon(t_i)$, where $A_i = \exp\left(-\frac{(t_i - t_{i-1})}{\tau} \right)$ represents the average autoregressive coefficient based on the average time spacing between observations, $\tau$ is the characteristic timescale, and $\epsilon(t_i)$ is a random error. The corresponding theoretical power spectrum of such a process is given by:

\begin{equation}\label{redfiteq}
    G_{rr}(f_i) = G_0 \frac{1 - A^2}{1 - 2 A \cos\left( \frac{\pi f_i}{f_{\text{Nyq}}} \right) + A^2}
\end{equation}

where $G_0$ denotes the mean spectral power, $f_i$ are the temporal frequencies, and $f_{\text{Nyq}}$ is the Nyquist frequency. To derive the red-noise-corrected power spectrum, we employed the REDFIT algorithm implemented in \texttt{R}\footnote{\url{https://rdrr.io/cran/dplR/man/redfit.html}}. The significance of the spectral features was determined using Equation~(\ref{redfiteq}). The analysis identified two distinct spectral peaks located at frequencies of \(0.003571 \pm 0.000700~\text{day}^{-1}\) and \(0.008747 \pm 0.000603~\text{day}^{-1}\), which correspond to periods of approximately 280 and 114 days, respectively. Both peaks exceed the 99\% confidence level (see Figure \ref{Fig-redfit}). The uncertainty in the peak frequency was estimated as the half-width at half-maximum (HWHM) from a Gaussian fit to the peak in the REDFIT spectrum .

\begin{figure*}
    \centering
    \includegraphics[width=0.95\textwidth]{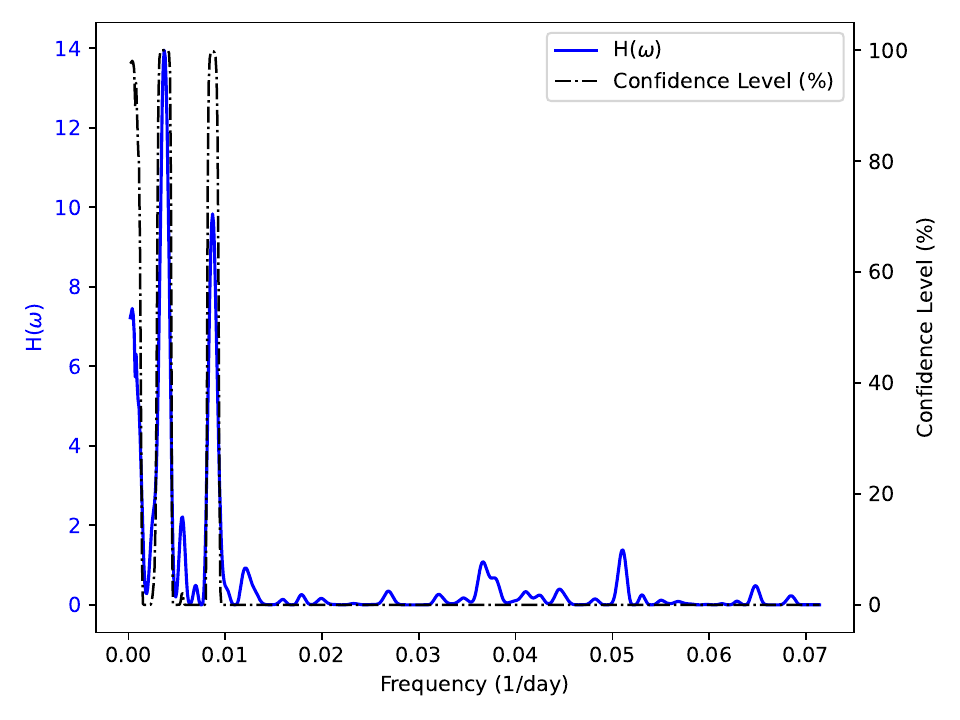}
    \caption{Modified periodogram $H(\omega)$ obtained using the DCDFT method, applied to the $\gamma$-ray light curve of the source. Two significant peaks are identified, corresponding to periods of approximately \(\sim 269.86\) days and \(\sim 114.54\) days. The periods uncertainties were estimated by fitting Gaussian functions to the periodogram peaks. The black dash-dotted curve shows the confidence level, which exceeds 99\% at both peaks, indicating strong statistical significance.}
    \label{Fig-DCDFT}    
\end{figure*}

\subsection{Date-compensated Discrete Fourier Transform (DCDFT)}\label{sec:dcdft}
The estimation of the power spectrum from unevenly sampled time series is a fundamental challenge in the search for quasi-periodic oscillations  in blazar light curves \citep[e.g.,][]{fan2007radio}. The standard discrete Fourier transform (DFT) method, when applied to such data, suffers from issues like frequency shifting and amplitude modulation, leading to unreliable detection of periodic components. These limitations are mitigated by the DCDFT method \citep{ferraz1981estimation, foster1995cleanest}, which fits the light curve using a least-squares regression model comprising sinusoidal and constant components, rather than relying solely on sinusoidal terms as in the classical DFT. This approach is particularly effective for low-frequency signals ($<0.02$\,d$^{-1}$), where the discrepancy introduced by uneven sampling can reach up to 5\%. 

In our analysis, the DCDFT was applied to the $\gamma$-ray light curve, resulting in the detection of two statistically significant periodic signals. We have used publicly available code.\footnote{\url{https://github.com/ilmimris/dcdft}}. The script implements the DCDFT based on the formalism of \citet{ferraz1981estimation}.
The modified periodogram $H(\omega)$ revealed peaks at \(0.003706 \pm 0.000556~\text{day}^{-1}\) and \(0.008731 \pm 0.000455~\text{day}^{-1}\), which correspond to periods of approximately 269.86 and 114.54 days, respectively. The confidence levels associated with both peaks exceeded 99\%, affirming their significance (see Figure \ref{Fig-DCDFT}). The frequency spacing in DCDFT remains uniform across the time series, irrespective of data gaps, making it well-suited for long-term blazar monitoring data with irregular sampling.

The results from all four approaches are consistent with each other, considering their uncertainties, summarized in Table \ref{tab:QPO_all}.

\begin{figure*}
    \centering
    \includegraphics[width=1.0\textwidth]{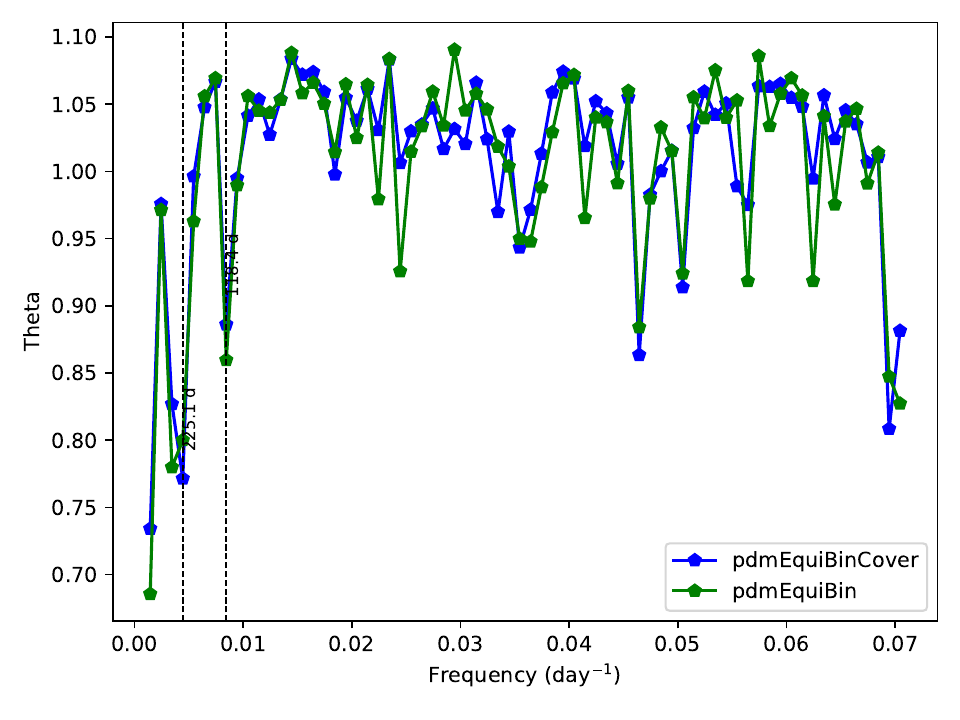}
    \caption{PDM analysis applied to the $\gamma$-ray light curve. The PDM $\theta$-statistic is plotted as a function of frequency. Two significant minima are observed, corresponding to periods of approximately 225.1 days and 118.4 days, indicating potential periodic behavior. Lower values of $\theta$ signify reduced phase dispersion, with the deepest minima representing the most coherent periodic signals.}
    \label{Fig-pdm}    
\end{figure*}

\subsection{The Phase Dispersion Minimization (PDM) analysis}

To complement our search for periodic signals, we applied the Phase Dispersion Minimization technique \citep{1978ApJ...224..953S}, which is particularly effective for detecting non-sinusoidal periodicities in unevenly sampled time-series data. The PDM method operates by folding the data over a range of trial periods and dividing the resulting phased light curve into bins. For each trial period, it computes the variance within each phase bin and compares it to the overall variance of the dataset. The resulting statistic, denoted as $\theta$, measures the phase dispersion: a lower $\theta$ value indicates a more coherent and periodic structure in the folded data. Unlike Fourier-based methods, PDM does not assume any functional form for the periodic signal, making it well-suited for identifying asymmetric or complex periodic behaviors. This flexibility is particularly advantageous in the context of blazar variability, where periodic signals may deviate significantly from simple sinusoidal profiles. We used the \texttt{PyAstronomy} package for this analysis.\footnote{\url{https://pyastronomy.readthedocs.io/en/latest/pyTimingDoc/pdm.html}}. Two distinct minima are identified at frequencies of \(0.00444~\text{day}^{-1}\) and \(0.00845~\text{day}^{-1}\), corresponding to approximate periods of 225.1 and 118.4 days, respectively, suggesting the presence of underlying periodic signals (see Figure~\ref{Fig-pdm}).

\begin{figure*}
    \centering
    \includegraphics[width=1.0\textwidth]{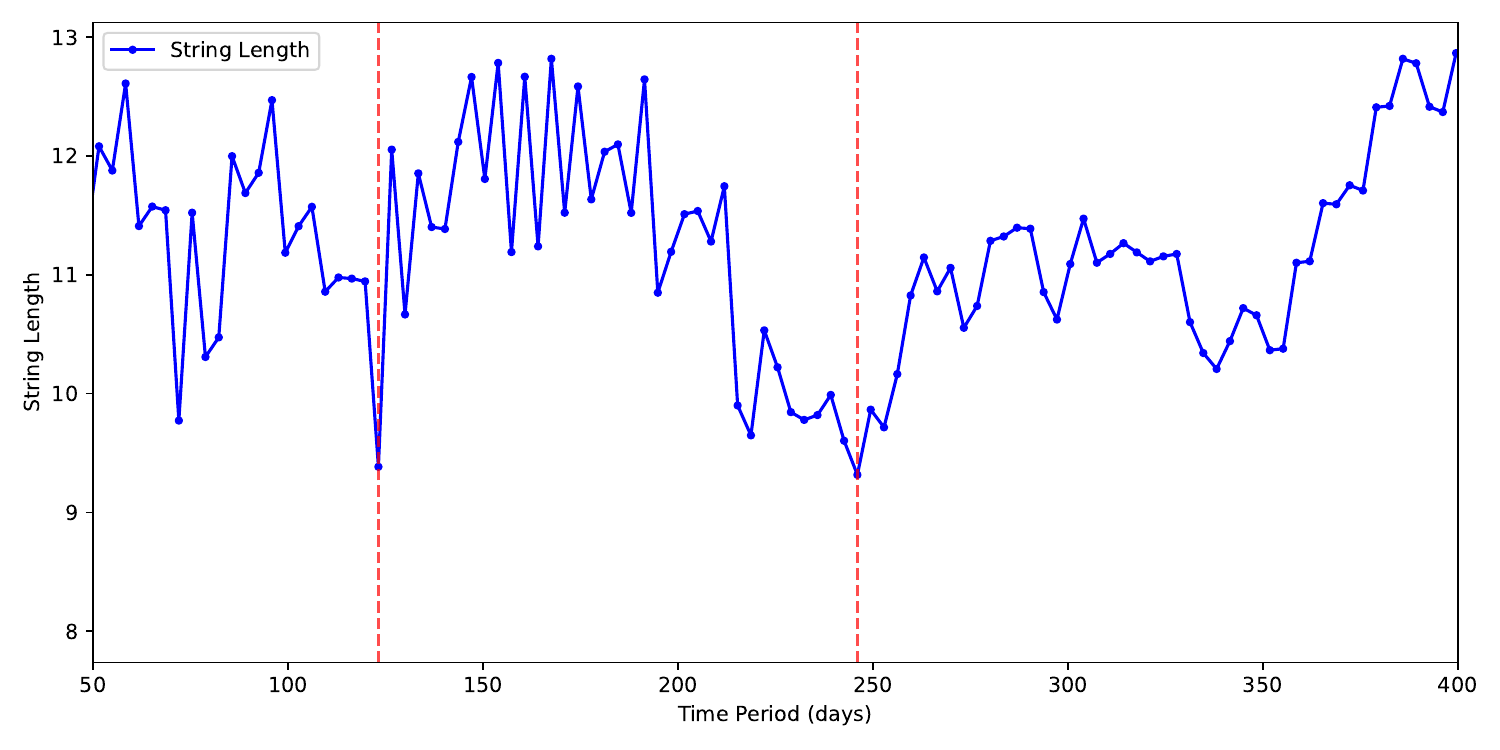}
    \caption{String-length analysis applied to the $\gamma$-ray light curve. Two prominent minima are observed at \(0.00407~\text{day}^{-1}\) and \(0.00812~\text{day}^{-1}\) which correspond to time periods of approximately 246.02 and 123.19 days,respectively, indicating candidate periodicities where the phase-folded data form the most ordered structures.}
    \label{Fig-slm}    
\end{figure*}

\subsection{The stringlength method}
To identify periodic signals in sparsely and irregularly sampled data, we applied the string-length method introduced by \citet{1983MNRAS.203..917D}. This technique is particularly well-suited for cases where the signal may be non-sinusoidal and the sampling cadence is uneven. The core idea is to fold the observational time series over a set of trial periods, sort the resulting phased data, and then compute the total "string length" by connecting adjacent points in phase space with straight-line segments. The period that minimizes this total length corresponds to the most coherent phase alignment, effectively identifying the true periodicity. Because it does not rely on sinusoidal fitting, the method remains sensitive to asymmetric or irregular periodic structures, which are common in blazar light curves and other astrophysical sources exhibiting complex variability.

The string-length method is a non-parametric, model-independent approach that requires no assumptions about the waveform of the periodic signal. This makes it especially valuable when exploring potential quasi-periodic oscillations  in blazars, where emission profiles can vary significantly from cycle to cycle. \citet{1983MNRAS.203..917D} demonstrated that the method performs reliably even when only a small number of high-precision observations are available, and proposed statistical criteria for assessing the validity of detected periods. In our analysis, we used this technique as a complementary tool alongside Fourier-based methods such as the LSP and DCDFT, enhancing our ability to detect both sinusoidal and non-sinusoidal periodic components in the $\gamma$-ray light curve. We used the \texttt{PyAstronomy} package for this analysis.\footnote{\url{https://pyastronomy.readthedocs.io/en/latest/pyTimingDoc/stringlength.html}}. Two prominent minima are found at frequencies of \(0.00407~\text{day}^{-1}\) and \(0.00812~\text{day}^{-1}\), corresponding to approximate periods of 246.02 and 123.19 days, respectively. These minima indicate candidate periodicities where the phase-folded light curve exhibits the most coherent structure (see Figure~\ref{Fig-slm}).\\

\begin{figure*}
    \centering
    \includegraphics[width=0.95\textwidth]{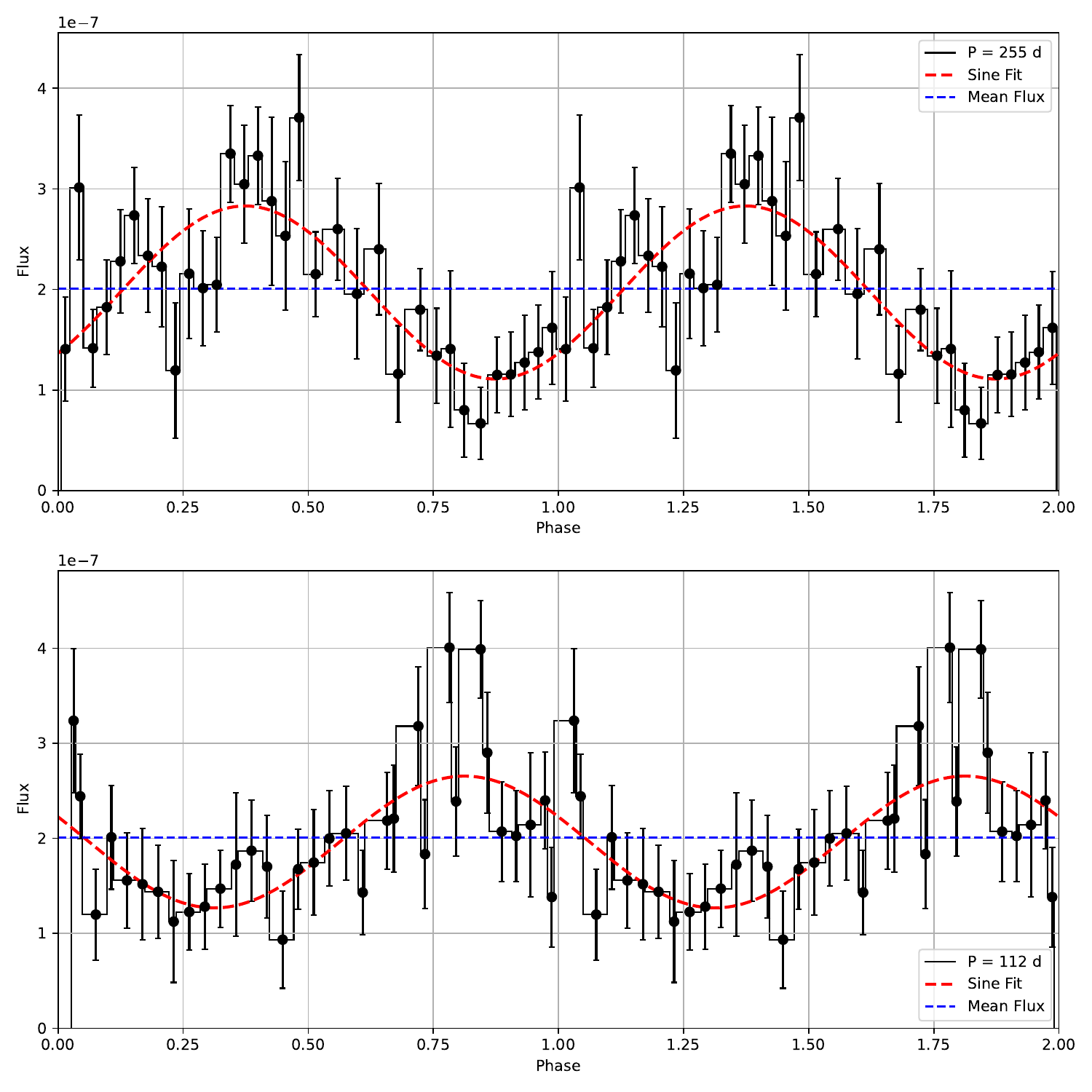}
    \caption{Phase-folded $\gamma$-ray light curves of the blazar PKS\,0805$-$07 for two significant periods identified from the LSP analysis. The top panel corresponds to the folding period \( P_1 = 254.8 \) days, and the bottom panel to \( P_2 = 112.1 \) days. In each panel, the observed flux is shown as black markers, overplotted with the best-fit sinusoidal model (solid blue line). The horizontal dashed line denotes the mean flux level. The folding over two full cycles (phase 0 to 2) highlights the periodic modulation of the $\gamma$-ray emission.}

    \label{Fig-phase}    
\end{figure*}

In this study, we present indication for two possible quasi-periodic oscillations in the $\gamma$-ray emission of the blazar PKS~0805$-$07, with characteristic periods of roughly 255 and 112 days. These signals surpass the $3\sigma$ significance threshold in multiple independent analyses, reinforcing their credibility.
To further confirm the presence of periodic signals, we constructed phase-folded $\gamma$-ray light curves using periods of approximately 255 days and 112 days, and fitted them with sinusoidal models (see Figure~\ref{Fig-phase}). The resulting fits provide additional support for the periodic nature of the emission from the blazar PKS~0805$-$07. For improved visual clarity, we display multiple periodic cycles in the phase-folded $\gamma$-ray light curves.

\begin{table*}
\setlength{\extrarowheight}{7pt}
\setlength{\tabcolsep}{7pt}
\centering
\caption{Summary of QPO frequencies detected using various methods. The uncertainties in the frequencies are quoted where available, and the local significance levels of the detections are provided in parentheses. The upper row shows results from spectral analysis methods: LSP, WWZ, REDFIT, and DCDFT. The lower row shows results from  PDM and String-Length . Frequencies are in units of $10^{-3}~\text{day}^{-1}$.}

\begin{tabular}{c c c c c}

\hline
4FGL Name  & LSP $\left(10^{-3}~\text{day}^{-1}\right)$ & WWZ $\left(10^{-3}~\text{day}^{-1}\right)$ & REDFIT $\left(10^{-3}~\text{day}^{-1}\right)$ & DCDFT $\left(10^{-3}~\text{day}^{-1}\right)$ \\
\hline
4FGL J0808.2$-$0751 
& 3.924$\pm$0.409 ($>99.9\%$) & 3.743$\pm$0.596 ($>99.5\%$) & 3.571$\pm$0.700 ($>99\%$) & 3.706$\pm$0.556 ($>99\%$)\\  
& 8.920$\pm$0.463 ($>99.9\%$) & 8.743$\pm$0.584 ($>99.5\%$) & 8.747$\pm$0.603 ($>99\%$) & 8.731$\pm$0.455 ($>99\%$)\\
\hline
\multicolumn{5}{c}{Additional Methods: PDM and String-Length } \\
\hline
4FGL J0808.2$-$0751 & PDM $\left(10^{-3}~\text{day}^{-1}\right)$ & String-Length $\left(10^{-3}~\text{day}^{-1}\right)$ &  & \\
\hline
 
& 4.440 (—) & 4.070 (—) & & \\
& 8.450 (—) & 8.120 (—) & & \\
\hline

\end{tabular}
\label{tab:QPO_all}
\end{table*}

\begin{figure*}
    \centering
    \includegraphics[width=0.95\textwidth]{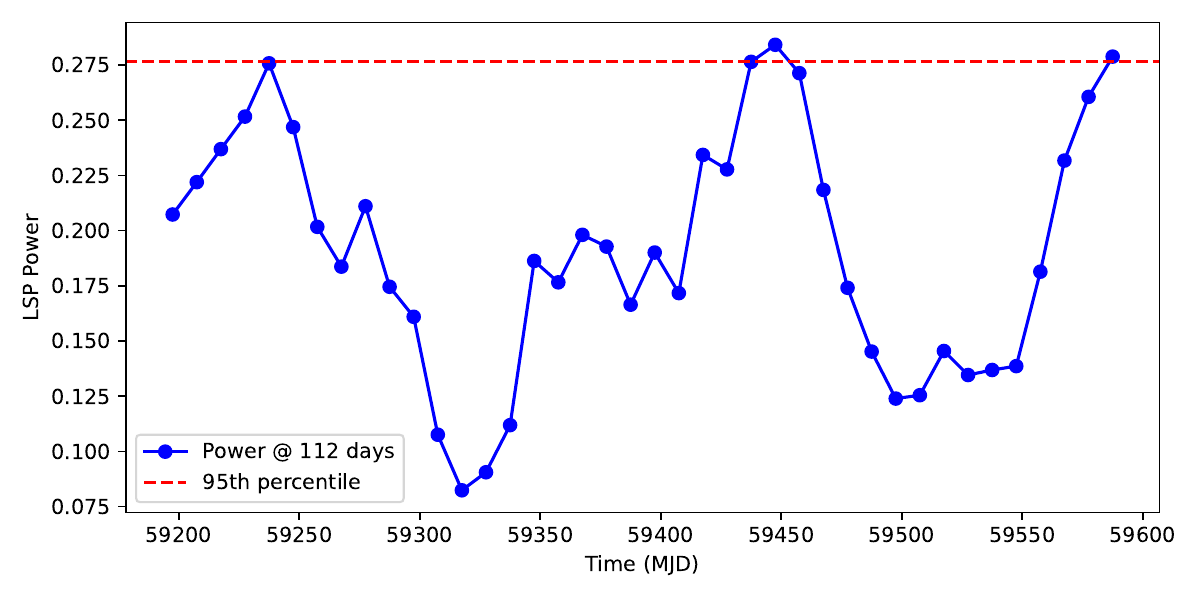}
    \includegraphics[width=0.95\textwidth]{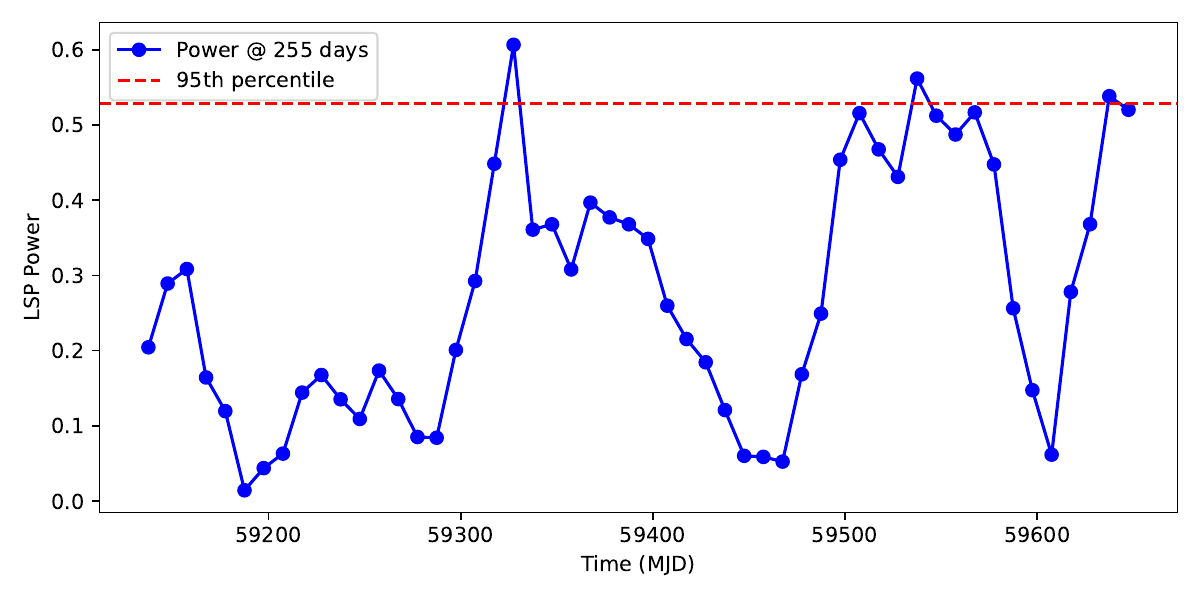}
    \caption{Windowed Lomb-Scargle power as a function of time at fixed periods of 112 and 255 days. This analysis employs a sliding window approach to track the temporal evolution of the quasi-periodic signals over the full observational baseline. The horizontal dashed line represents the 95th percentile significance threshold. Intervals where the power exceeds this threshold indicate that the QPOs are intermittent in nature and do not persist uniformly throughout the light curve.
}

    \label{Fig-window}    
\end{figure*}

\section{Temporal Evolution of the QPO}
\label{evo}
To investigate the temporal stability of the QPOs detected near 112 and 255 days, we performed a windowed LSP analysis. In this approach, the light curve was segmented using a sliding time window, and the LSP power at the fixed periods of 112 and 255 days was computed within each segment. The resulting time series of LSP power, shown in Figure~\ref{Fig-window}, reveals the variable nature of the signal strength over the course of the observation. Notably, the power exceeds the 95th percentile threshold during multiple intervals, indicating that the QPOs are not persistent throughout the entire light curve but instead occur in localized episodes. This behavior is characteristic of transient or evolving QPOs and suggests that the underlying mechanism may be episodic in nature, possibly linked to dynamic processes in the jet or accretion environment of the blazar.

\begin{figure*}
    \centering
    \includegraphics[width=0.9\textwidth]{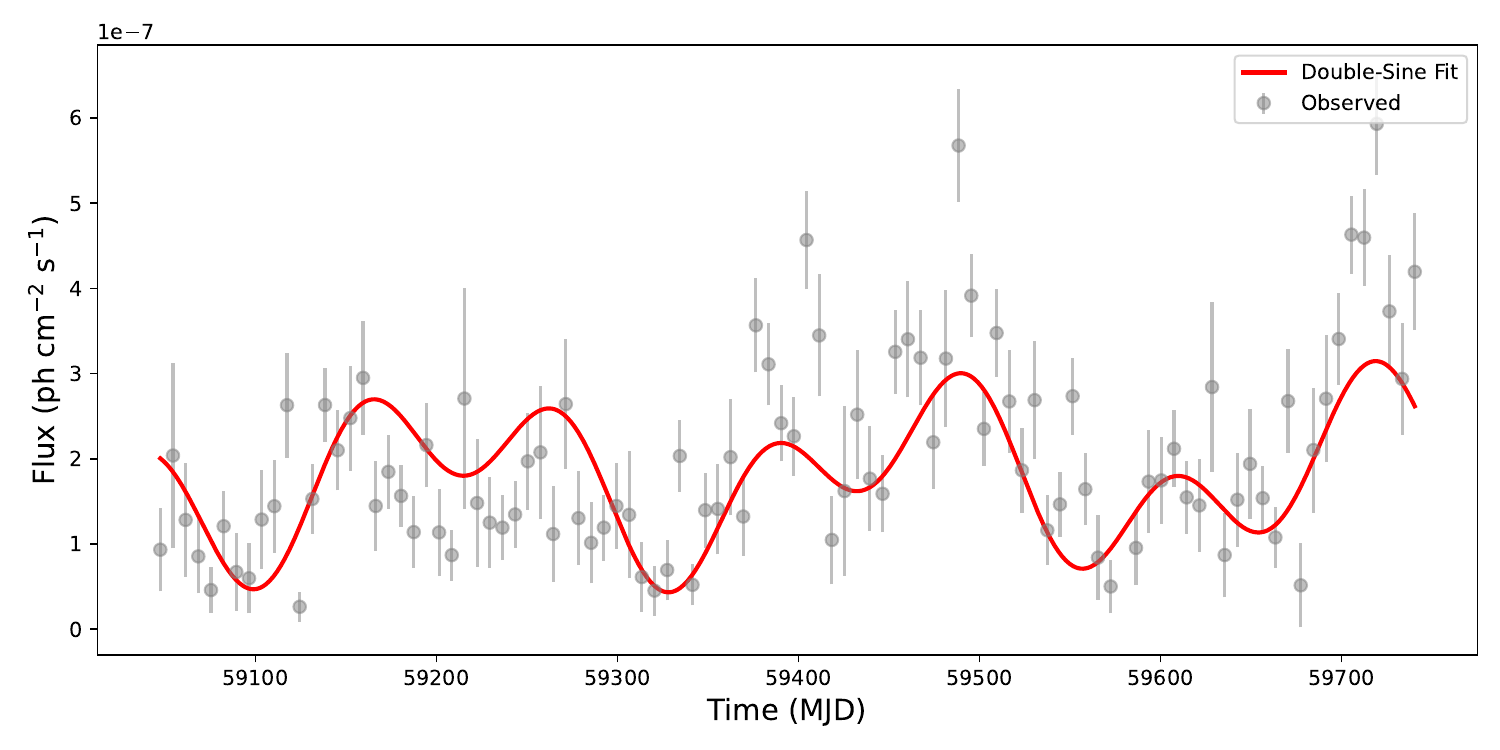}
    \caption{
    Double-sine function fitted to the observed $\gamma$-ray light curve of PKS\,0805-07. The model incorporates two sinusoidal components with periods of 254.1 days and 112.8 days, respectively. The observed flux (black points) is overplotted with the combined model (solid red line), illustrating how the superposition of the two periodic signals reproduces the overall modulation pattern in the light curve.
    }
    \label{fig:doubleSine}
\end{figure*}

\section{Modeling the Quasi-Periodic Variability with a Double Sine Function}
\label{sine_fit}
To further investigate the quasi-periodic behavior observed in the $\gamma$-ray light curve of PKS\,0805-07, we fitted a double-sine model incorporating two periodic components. The best-fit model includes sinusoidal variations with periods of approximately 254.8 and 112.1 days. As shown in Figure~\ref{fig:doubleSine}, the combined model (red curve) captures the modulation pattern in the observed light curve (black points), suggesting that the variability may arise from the superposition of two independent periodic processes. This dual-periodic structure supports the presence of multiple QPO signatures, possibly associated with different emission zones or physical mechanisms such as jet precession and internal shocks. The successful fit highlights the relevance of multi-component models in describing the complex temporal behavior of blazars. We model the observed flux as the sum of two sinusoidal components:

\begin{equation}
F(t) = A_1 \sin\left( \frac{2\pi t}{P_1} + \phi_1 \right) + A_2 \sin\left( \frac{2\pi t}{P_2} + \phi_2 \right) + C,
\end{equation}

where \( F(t) \) is the flux as a function of time \( t \) (in MJD), \( A_1 \) and \( A_2 \) are the amplitudes, \( \phi_1 \) and \( \phi_2 \) are the phases, and \( C \) is a constant offset representing the mean flux level. The two periods are fixed at \( P_1 = 254.8 \) days and \( P_2 = 112.1 \) days, corresponding to the dominant QPOs detected in the power spectrum.

The best-fit parameters obtained from the double-sine fitting are:
\[
A_1 = (6.98 \pm 0.66) \times 10^{-8}, \quad \phi_1 = -0.83 \pm 0.10,
\]
\[
A_2 = (-6.72 \pm 0.70) \times 10^{-8}, \quad \phi_2 = 0.15 \pm 0.10,
\]
\[
C = (1.78 \pm 0.05) \times 10^{-7}.
\]

The double-sine model supports the presence of two independent modulations in the $\gamma$-ray light curve of PKS\,0805-07. The longer-term component, with a period of approximately 254.8 days, may reflect a quasi-periodic or geometric origin, such as jet precession or wobbling of the emission axis. In contrast, the shorter-term modulation near 112.1 days is likely associated with intrinsic emission processes within the jet or potential dynamical interactions, such as those arising from a binary supermassive black hole system. The superposition of these two signals provides a natural explanation for the complex variability observed in the source.
The folded light curve (see Figure \ref{Fig-phase}) at \( 254.8 \) days exhibits a positive, symmetric sinusoidal variation, consistent with the positive amplitude and right-shifted phase derived from the fit. In contrast, the 112-day folded curve shows an inverted sinusoidal pattern, in agreement with the negative amplitude and slightly left-shifted phase. The strong visual agreement between the model predictions and the folded light curves reinforces the interpretation that the observed variability is shaped by the coherent superposition of two periodic signals.

\begin{figure*}
    \centering
    \includegraphics[width=0.85\textwidth]{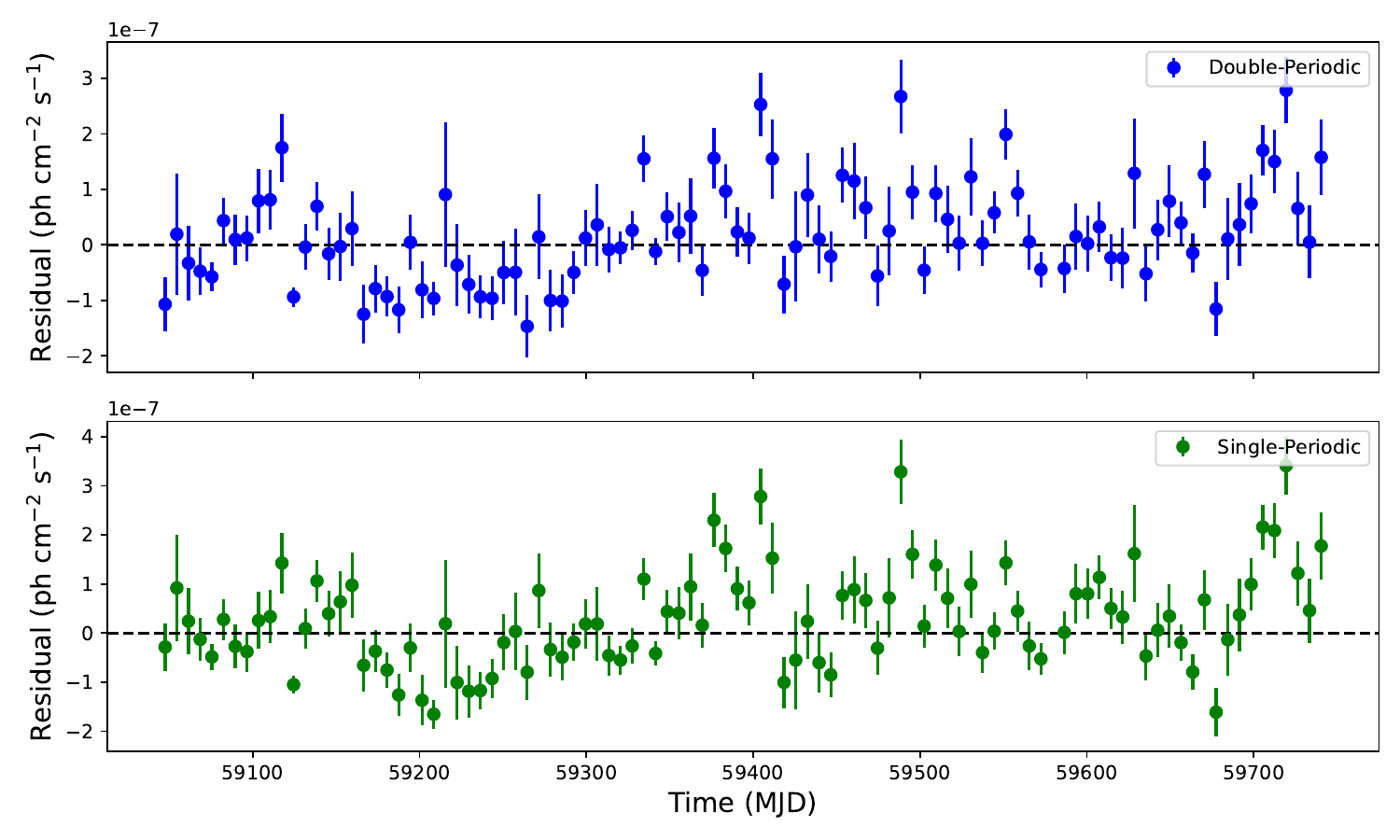}
    \caption{
    Residuals from model fits to the $\gamma$-ray light curve of PKS\,0805-07. 
    \textbf{Top panel:} residuals from the double-periodic model incorporating both 254.8-day and 112.1-day components. 
    \textbf{Bottom panel:} residuals from the single-periodic model using only the 254.8-day component. 
    The residuals from the double-sine fit are smaller in amplitude and show no systematic trends, 
    indicating a significantly improved fit compared to the single-component model.
    }
    \label{fig:residuals}
\end{figure*}

To evaluate whether the addition of a second periodic component significantly improves the fit to the $\gamma$-ray light curve of PKS\,0805-07, we compared three models using the Akaike Information Criterion (AIC) and Bayesian Information Criterion (BIC): a null (constant) model, a single-sine model with a period of 254.8 days, and a double-sine model incorporating both the 254.8-day and 112.1-day components. The AIC values for the null, single-periodic, and double-periodic models were 186.65, 147.42, and 125.99, respectively. The corresponding BIC values were 189.24, 155.21, and 138.97. Both AIC and BIC show a substantial improvement when moving from the single- to the double-periodic model, indicating that the inclusion of the second periodic signal yields a significantly better fit despite the additional model complexity. Residual plots further confirm this conclusion by showing reduced scatter and no systematic deviation for the double-sine fit (see Figure~\ref{fig:residuals}). These findings statistically reinforce the presence of two distinct quasi-periodic components in the source variability.

\begin{figure*}
    \centering
    \includegraphics[width=0.9\textwidth]{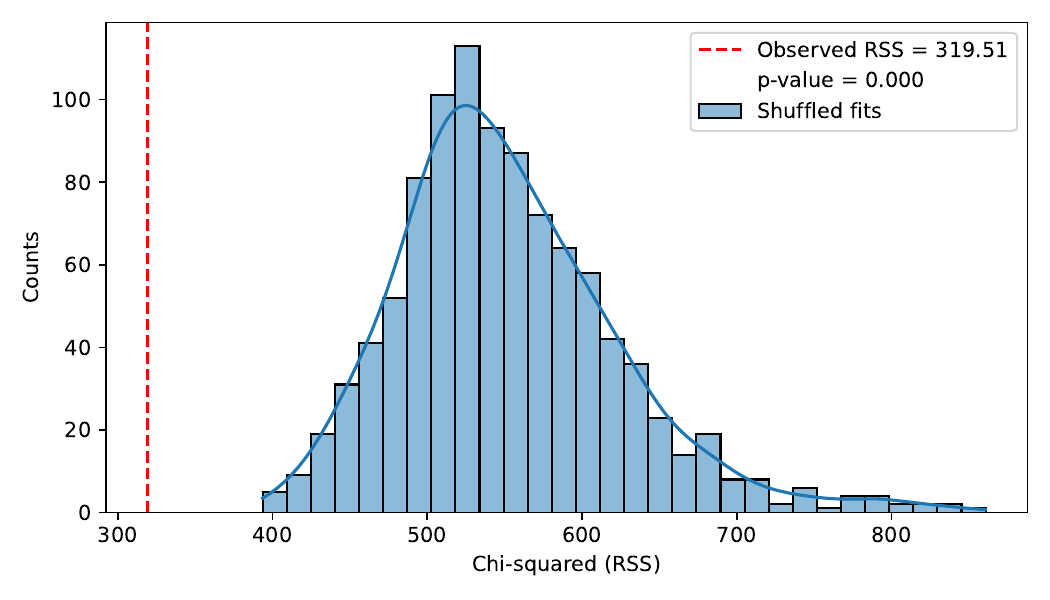}
    \caption{
    Monte Carlo significance test for the double-periodic model. The histogram shows the distribution of RSS values obtained by fitting 1000 flux-shuffled light curves with the same double-sine model. The vertical dashed line marks the RSS of the original, unshuffled data (\( \text{RSS} = 319.51 \)). The observed value lies in the extreme left tail of the distribution, with a computed p-value of \( < 0.001 \), indicating that the observed fit is highly unlikely to result from random fluctuations and is therefore statistically significant.
    }
    \label{fig:mc_rss}
\end{figure*}

To assess the statistical significance of the double-sine model, we conducted a Monte Carlo simulation test. The flux values were randomly shuffled 1000 times while keeping the time array fixed, and each shuffled light curve was fitted using the same model with fixed periods of 254.8 and 112.1 days. For each trial, we calculated the residual sum of squares (RSS), thereby constructing a null distribution under the hypothesis that the observed periodicity arises purely by chance. The RSS value for the original data was found to be significantly lower than that of the vast majority of shuffled realizations (see Figure~\ref{fig:mc_rss}), yielding a  p-value of \( < 0.001 \). This result confirms that the detected periodic components are statistically significant and unlikely to be artifacts of random fluctuations.
To further test the stability of the detected periodicities, we re-fitted the double-sine model by allowing the periods to vary as free parameters, rather than fixing them at 254.8 and 112.1 days. The resulting best-fit parameters and their uncertainties are:
\begin{align*}
A_1 &= (6.98 \pm 0.66) \times 10^{-8}, \quad
A_2 = (6.72 \pm 0.70) \times 10^{-8}, \\
C   &= (1.78 \pm 0.05) \times 10^{-7}, \\
P_1 &= 254.16 \pm 5.49 \, \text{days}, \quad
P_2 = 112.20 \pm 1.02 \, \text{days}.
\end{align*}

These periods are consistent with the peaks found in the Lomb–Scargle periodogram, showing that our model recovers the same main signals. This supports the idea that the detected periodicities are real and not just artifacts of the fitting process.

\section{Amplitude-Modulated Jet Precession Model}\label{beat}
The double sine model fit presented in Section \ref{sine_fit} revealed two significant periodic components with comparable amplitudes, suggesting the possibility of a beating phenomenon arising from their interference. The nearly equal modulation strengths of these oscillatory terms motivate a physical interpretation in terms of a geometric origin rather than an intrinsic flux variation. To explore this, we investigated a scenario involving the precession of a relativistic jet with small-amplitude angular modulations. In this framework,  we consider a scenario in which the relativistic jet undergoes slow precession, with its orientation varying periodically in both azimuthal and polar directions. This variation causes the spiral jet structure to tighten or loosen, modulating the viewing angle and hence the observed flux.\\

We assume the observer lies in the \( yz \)-plane, such that the observer's unit vector is:
\( \vec{O}=(0, \sin i, \cos i) \approx (0, i, 1-i^{2}/2) \) for \( i \ll 1 \).
where \( i \) is the inclination angle, assumed to be small.
The jet direction vector is given by 
\( \vec{J}(t)=(\sin\theta(t)\sin\phi(t), \sin\theta(t)\cos\phi(t), \cos\theta(t)) \), 
which under the small-angle approximation becomes 
\( \vec{J}(t) \approx (\theta(t)\sin\phi(t), \theta(t)\cos\phi(t), 1-\theta(t)^2/2) \).
 The jet azimuthal and polar angles evolve with time as:
\begin{equation}\label{az_po}
\phi(t) = 2\pi f_{1} t + \phi_1, \quad
\theta(t) = \theta_0 \cos(2\pi f_{2} t + \phi_2),
\end{equation}
with \( f_1 \) and \( f_2 \) being the angular frequencies of azimuthal and polar motion respectively, and \( \phi_1 \), \( \phi_2 \) their respective phases. The angle \( \alpha(t) \) between the jet and the observer is then given by:
\begin{equation}
\cos\alpha(t) = \vec{O} \cdot \vec{J}(t) \approx 1 + i\theta(t)\cos\phi(t) - \frac{i^2}{2} - \frac{\theta^2(t)}{2} + \mathcal{O}(i^2\theta^2)
\end{equation}
Neglecting \hspace{0.02cm} $\mathcal{O}(i^2\theta^2)$ term, the relativistic Doppler factor (for small $\epsilon$), depends on this angle as:
\begin{equation}
\delta(t) = \frac{1}{\Gamma(1 - \beta \cos\alpha(t))}
\approx \frac{1}{\Gamma(1 - \beta)} \left(1 + \frac{\epsilon}{1 - \beta} \right) 
\end{equation}
where \( \Gamma \) is the bulk Lorentz factor, \( \beta = v/c \) and $\epsilon = \beta\left(i\theta\cos\phi - \frac{i^2}{2} - \frac{\theta^2}{2} \right)$ \\\\
Since $\beta \approx 1 - \frac{1}{2\Gamma^2}$, then $1 - \beta \approx \frac{1}{2\Gamma^2}$ and $\beta \Gamma{^2} \approx \Gamma^{2} $( for $\Gamma>1$), so the flux $(F(t))$ is given by:
\begin{equation}\label{eq_1}
F(t) \propto \delta(t)^p \approx (2\Gamma)^p \left(1 + 2\Gamma^2 p \epsilon \right)
\end{equation}
\\\\
Substitute $\epsilon$ in \ref{eq_1}, we get (see Appendix \ref{appen} )
\begin{align}
F(t) \propto F_0 \Bigg[1 + \frac{dF_1}{2} \big( \cos((2\pi f_-)t + \phi_-)  \nonumber \\
+ \cos((2\pi f_+)t + \phi_+) \big) \nonumber \\
- \frac{dF_2}{2} \cos(4\pi f_2 t + 2\phi_2) \Bigg]
\end{align}\\
where,
\begin{equation}
f_\pm = f_1 \pm f_2, \quad
\phi_\pm = \phi_1 \pm \phi_2.
\end{equation}

This model naturally explains the presence of two frequency components (\(f_-\) and \(f_+\) ) in the observed light curve as arising from the precessing motion of a relativistic jet whose orientation is modulated over time, leading to amplitude-modulated Doppler boosting. The term proportional to \(\theta^2\) in the flux expansion of Equation \ref{eq_a2} introduces a cosine modulation at frequency \(2 f_2\). This implies that if the \(\theta^2\) contribution is retained, a weak signal at frequency \(2 f_1\) or \(2 f_2\) should be present in the power spectrum. However, a careful search using LSP, WWZ, and other time-series analysis methods reveals no significant periodicities near \(2 f_1\) or \(2 f_2\), suggesting that the amplitude of this harmonic component is likely below the detection threshold in the current data.\\\\
For \( \theta \ll i \), the equation reduces to:
\begin{align}\label{eq_3}
F(t) &\propto  F_0 \left[1 + \frac{dF}{2} \left( \cos((2\pi f_-)t + \phi_-) + \cos((2\pi f_+)t + \phi_+) \right) \right]
\end{align}\\
Where, 
$F_0 = 2^p \Gamma^p (1 - \Gamma^2 p i^2)$ and 
$dF = \frac{2\Gamma^2 p i \theta_0}{1 - \Gamma^2 p i^2}$ \\\\
Under the approximation $i^2 \ll 1/\Gamma^2$, we may further simplify:
\[
1 - \Gamma^2 p i^2 \approx 1, \quad \Rightarrow  dF \approx 2 \Gamma^2 p i \theta_0
\]

\begin{figure*}
    \centering
    \includegraphics[width=0.9\textwidth]{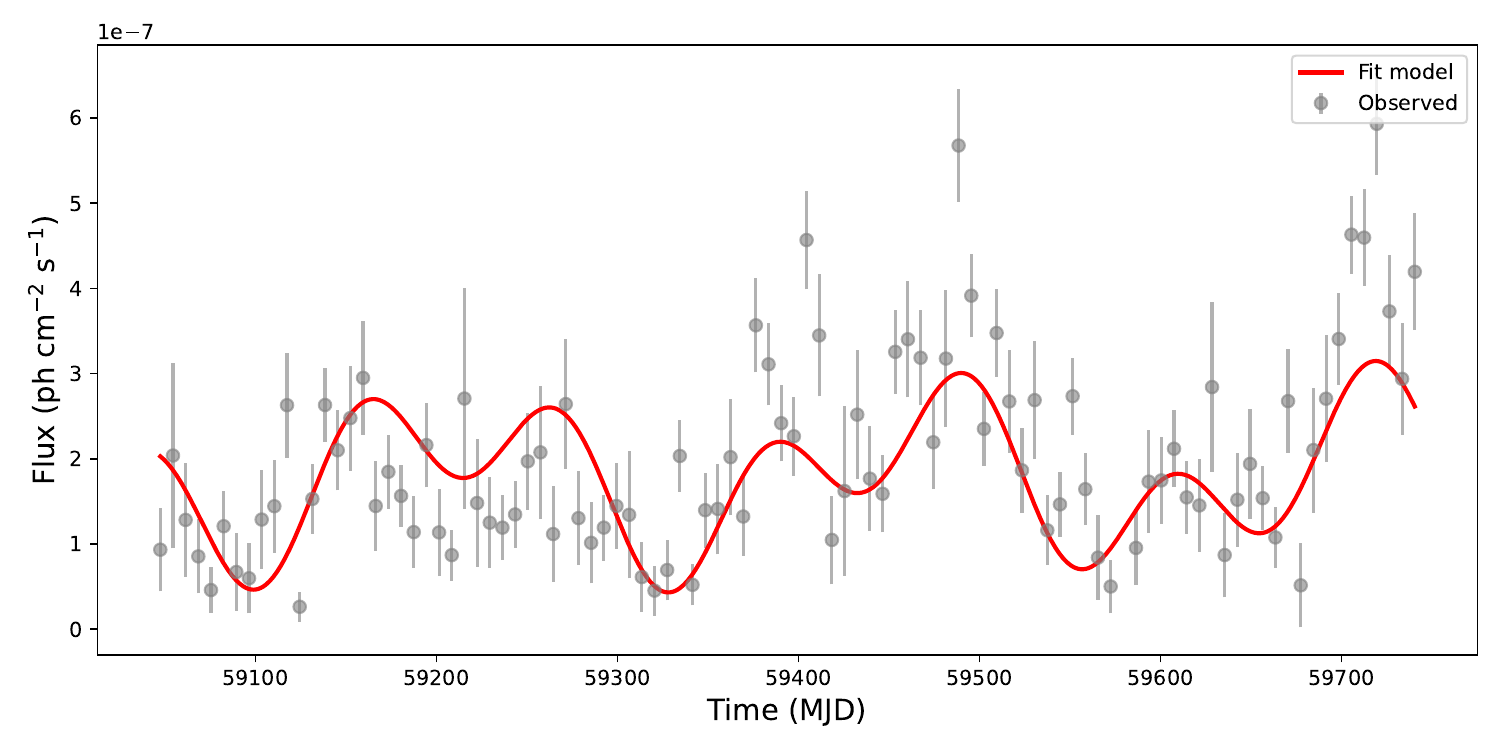}
    \caption{
    Fit of the amplitude-modulated jet precession model to the observed $\gamma$-ray light curve. The model captures the beat-like modulation due to jet precession and viewing angle variation.
    }
    \label{fig:cosine_expansion_fit}
\end{figure*}

To quantitatively test the applicability of the amplitude-modulated Doppler boosting model described by Equation \ref{eq_3}, we fitted it directly to the observed $\gamma$-ray light curve. Using a non-linear least-squares method, we obtained the best-fit parameters: $F_0=(1.777 \pm 0.005) \times 10^{-7}\,\mathrm{ph\,cm^{-2}\,s^{-1}} $, $dF=0.772 \pm 0.049$, with corresponding frequencies \( f_1=(25.03 \pm 0.0038) \times 10^{-4} \mathrm{day^{-1}} \) and 
\( f_2=(64.22 \pm 0.0032) \times 10^{-4} \mathrm{day^{-1}} \). The model curve successfully reproduces the beating pattern observed in the light curve (see Figure~\ref{fig:cosine_expansion_fit}). These results support the physical plausibility of amplitude-modulated jet precession as a mechanism for generating the observed dual periodicities in PKS\,0805-07.
Substituting the given values: $ dF=0.772 $,  $ \Gamma=10 $, and $p=5 $, $i\theta_0=7.72 \times 10^{-4}$. Parameter sets satisfying \(\theta_0 \ll i \ll 1/(2\Gamma)\), with \(p = 5\), and yielding \(dF \approx 0.77\) are shown in Table \ref{tab_p}.
\begin{table}
\centering
\begin{tabular}{|c|c|c|c|c|c|}
\hline
\(\Gamma\) & \(i\) & \(\theta_0\) & \(1/(2\Gamma)\)&\(\theta_0/i\) & \(2\Gamma i\) \\
\hline
20 & 0.01962 &0.00981& 0.025& 0.5 & 0.7848 \\
25 & 0.01555 & 0.00792& 0.020& 0.5 & 0.7775  \\

30 & 0.01308 & 0.00654&0.0167 & 0.5 & 0.7848  \\
35 & 0.01120 & 0.00561&0.0142 & 0.5 & 0.7840 \\

40 & 0.00976 & 0.00493&0.0125 & 0.5 & 0.7808  \\
\hline
\end{tabular}
\caption{Parameter sets satisfying \(\theta_0 \ll i \ll 1/(2\Gamma)\), with \(p = 5\), and yielding \(dF \approx 0.77\).}
\label{tab_p}
\end{table}

It is important to note that the fractional variability amplitude \( dF \), which quantifies the modulation depth of the observed flux, depends directly on the bulk Lorentz factor \( \Gamma \) of the jet. Specifically, under geometric modulation models, \( dF \propto \Gamma^2 \). This suggests a natural explanation for the absence of QPOs in low-flux states: during such states, the jet may be in a slower, less relativistic phase with a reduced Lorentz factor. As a consequence, the resulting Doppler boosting—and thus the flux modulation amplitude—would be diminished. The reduced value of \( \Gamma \) leads to a smaller \( dF \), potentially suppressing QPO signatures below detectability thresholds. This scenario aligns with the idea that geometric effects, such as jet precession or wobbling, can only imprint strong observational signatures when the relativistic beaming is sufficiently high. In our case the QPO period (MJD 59047.5-59740.5) includes the majority of active period (MJD
59370-59965).

The precession and modulations are occurring in the lab frame.
The angular frequencies $f_-$ and $f_+$ correspond to the observed values. The observed frequencies are related to the intrinsic ones by $f_{\text{obs}} = \frac{f_{\text{int}}}{1 - \beta \cos\alpha(t)}$, which, under the approximation \(\cos\alpha(t) \approx 1\) and \(\beta \approx 1 - \frac{1}{2\Gamma^2}\), simplifies to
$f_{\text{obs}} \approx 2\Gamma^2 f_{\text{int}}$. For $\Gamma=10$, $f_{\text{obs}} \approx 200 \hspace{0.02cm} f_{\text{int}}$. That means the  intrinsic time period ($T_{int}$) $\approx$ 200 times the observed time period ($T_{obs}$). For $T_{obs}=255 \hspace{0.02cm}d$, $T_{int}$ $\approx 51000 \hspace{0.02cm}d$.
The time-dependent angle $\alpha(t)$ introduces fluctuations in $\delta(t)$ through small variations in $\theta(t)$. However, under the assumption $i \gg \theta$, the viewing angle is dominated by the constant inclination $i$, and the modulation due to $\theta(t)$ becomes negligible. As a result, the Doppler factor can be treated as approximately constant in time, and both $f_1$ and $f_2$ are scaled by a fixed factor. Therefore, the $\theta$-modulation affects the amplitude of the observed flux but not the observed timescales or frequencies of variability.\\

The period of Lense–Thirring (LT) precession can be estimated as \citep{2020ApJ...891..120B}:
\begin{equation}
t_{\mathrm{LT}} = 0.18 \left( \frac{1}{a_*} \right) \left( \frac{M}{10^9\,M_\odot} \right) \left( \frac{r}{r_g} \right)^3 \ \text{days}
\end{equation}\\\\
where \( a_* \) is the dimensionless spin parameter, \( M \) is the mass of the black hole, and \( r \) is the radial distance of the  base of the jet where LT precession takes place, expressed in gravitational radii (\( r_g = GM/c^2 \)). For \( M = 10^9 M_\odot \), \( a_* = 0.9 \) and \( T_{\text{int}} (1/f_{1}) = 80{,}000 \, d = t_{\text{LT}} \), we get $\frac{r}{r_g} \approx 73.65$. Therefore, to match  precession period of \(400\) days, the intrinsic LT period should be \(\sim80{,}000\) days (219 years). This corresponds \( r \) approximately \(73.65\,r_g\) for a \(10^9\,M_\odot\) black hole with spin \(a_* = 0.9\). The 219 year time scale is consistent with expectations from Lense–Thirring precession \citep{2007Ap&SS.309..271R}, arising from a characteristic radius of $\sim 73 \hspace{0.05cm}r_g$ around a supermassive black hole, which may be interpreted as the base radius of the jet ($R_B$). If instead the characteristic radius were closer to $10\hspace{0.05cm} r_g$, the Lense–Thirring precession time scale would be significantly smaller -- approximately 200 days. We note that these estimates depend sensitively on the assumed values of the bulk Lorentz factor and the black hole mass. 

For a conical jet with an opening angle $\theta_{j}$, the transverse radius of the  jet ($R_C$)  is given by:
\begin{equation}
    R_C \approx R_B + \theta_{j} \cdot l \approx \theta_{j} \cdot l \hspace{0.3cm}(R_{B}\ll \theta_{j} \cdot l)
\end{equation}
where $l$ is the distance of emission region from the base of the jet.
The shortest variability timescale observed in $\gamma$-rays, as reported in our previous work \citep{2024ApJ...977..111A}, is $t_{\text{var}} = (2.80 \pm 0.77)$ days. This places an upper limit on the size of the emission region ($S_B$):
\begin{equation}
    S_B \lesssim \frac{c\, t_{\text{var}}\, \delta}{1 + z},
\end{equation}
Substituting $\delta = 20$ and $z = 1.837$ , we obtain $S_B \lesssim (5.11 \pm 1.41) \times 10^{16}\,\text{cm}$.
Since the emission region should lie within the jet cone, we require \( S_B < R_C \). Taking the opening angle as \( \theta_{j}= 0.01555~\text{radian} \) (for \( \Gamma = 25 \), see Table~\ref{tab_p}), this condition implies:
\begin{equation}
    R_C \approx \theta_{j} \cdot l > S_B \quad \Rightarrow \quad l > \frac{S_B}{\theta_{j}} > 1 \, \text{parsec}.
\end{equation}
Thus, the emission must occur at a distance \( l > 1\,\text{parsec} \). This is consistent with our previous broadband SED modeling \citep{2024ApJ...977..111A}, which demonstrated that external Compton (EC) scattering of infrared photons from the dusty torus provides a good fit to the multiwavelength data, placing the emission zone at parsec-scale distances.

\section{Summary and Discussion}
\label{sum}

In this study, we conducted a detailed time-series analysis of the $\gamma$-ray light curve of the FSRQ PKS\,0805-07 using data from the Fermi-LAT. Our goal was to identify and characterize any QPOs that may be present. Through the application of a suite of complementary time-domain techniques-including the LSP, WWZ, REDFIT (AR(1) noise-corrected spectrum), DCDFT, PDM, and the string-length method-we consistently detected two dominant periodic signals at characteristic timescales of approximately 255 days and 112 days.
Each method independently identified these periodicities with local significance exceeding 99\%, and the associated uncertainties in the detected frequencies were estimated using Gaussian fitting to the periodogram peaks. Furthermore, we constructed phase-folded light curves at both periods and observed coherent modulations consistent with sinusoidal behavior, validating the presence of periodic signals in the observed variability. A double-sine model was then fitted to the light curve, yielding amplitudes, phases, and mean flux values that align well with the phase-folded structures. The signs and relative amplitudes of the fitted components were consistent with the symmetry and phase alignment of the observed phase-folded modulation patterns, reinforcing the reliability of the identified QPOs.

To quantitatively evaluate the significance of the two--component periodic model, we performed model selection using AIC and BIC. Both criteria favored the double-periodic model over single-periodic and null (constant flux) alternatives, indicating a statistically superior fit. In addition, a Monte Carlo simulation test was conducted by randomly shuffling the flux values 1000 times and fitting each synthetic light curve with the same model. The resulting null distribution of chi-squared residuals demonstrated that the observed fit lies in the extreme left tail, with a p-value $< 0.001$. This confirms that the likelihood of obtaining such a good fit by chance is extremely low. To further examine the stability and reliability of the detected periodicities, we re-fitted the double-sine model by treating the periods as free parameters instead of fixing them at 254.8 and 112.1 days. The best-fit periods obtained from this approach are consistent with the peak frequencies identified in the Lomb–Scargle periodogram, indicating that the model independently recovers the same dominant signals. This agreement reinforces the interpretation that the detected modulations are intrinsic to the source rather than artifacts of model assumptions or fitting constraints.
Taken together, these results offer compelling indication for the presence of two statistically significant QPOs in the light curve of PKS\,0805-07. The combination of multiple analytical methods, phase-folded validation, statistical model selection, and Monte Carlo-based significance testing establishes a rigorous and reproducible framework for QPO detection in blazars.

The detection of two statistically significant periods of approximately 255 days and 112 days in the $\gamma$-ray light curve of PKS\,0805-07 is a noteworthy finding. Such dual QPOs are rare in blazars and prompt an exploration of their underlying mechanisms.  Since PKS\,0805-07 is located at a relatively high redshift ($z = 1.837$), it is therefore a potential candidate for hosting a binary SMBH system \citep{2017MNRAS.471.3036P,2009ApJ...703L..86V}.  A plausible explanation for QPOs is the presence of such a binary; however, true binary SMBH orbital periods are expected to span several years  \citep{2007Ap&SS.309..271R}, making the observed timescales ($\sim$ 255 d and $\sim$ 112 d) far too short to be explained by orbital motion alone \citep{2008Natur.452..851V}. Applying this framework to PKS\,0805-07 would require an unrealistically compact, massive system or extremely close separations-that are highly improbable given known SMBH mass-separation constraints \citep{2013A&A...552A..11R}. 

Another possibility is that a companion SMBH induces precession of the accretion disk and jet, causing periodic modulation of the jet emission on timescales shorter than the binary's orbital period. In OJ 287, for example, the $\sim 12$ yr optical QPO arises when a secondary SMBH impacts the primary's accretion disk \citep{1996ApJ...460..207L,1997ApJ...484..180S,2006ApJ...643L...9V}. At the same time, Lense-Thirring torques on the tilted disk can cause the jet axis to precess over timescales ranging from a few months up to a year or more, depending on parameters like spin and disk radius. By analogy, if PKS\,0805-07  primary black hole is rapidly spinning and its inner disk is tilted, Lense-Thirring precession could happen on roughly a 400 d cycle-falling comfortably within the “months-to-year” range. A second, shorter-timescale process (for instance, a disk-instability cycle or nodal precession of an inner warped disk region) could then produce the shorter periodicity.  Nonetheless, no high-resolution VLBI monitoring or polarization studies currently reveal systematic jet-position-angle swings or periodic polarization rotations in PKS\,0805-07.
Another mechanism involves periodic internal shocks propagating down the relativistic flow. If the central engine launches plasma blobs at quasi-regular intervals (for instance, due to disk oscillations, magnetic-reconnection cycles), each disturbance can steepen into a shock that enhances particle acceleration and radiative efficiency. Different regions or processes could then operate on different timescales, leading to multiple QPOs. For example, if an accretion-disk instability cycles every $\sim 255$ d thereby  modulating conditions at the jet base and a secondary instability in the corona or disk wind produces shocks every $\sim 112$ d further downstream, two distinct $\gamma$-ray QPOs would result. In the blazar S4 0954+658, two transient $\gamma$-ray QPOs of $\sim 66$ d and $\sim 210$ d were detected and attributed to a plasma blob moving helically inside the jet, although shocks and multiple emission zones were also considered plausible \citep{2023ApJ...949...39G}.

Interestingly, we noted that the periodic emission is supported by the double-sine fit analysis, which reveals two periodic components with longer-period component ($\sim$255 days) showing a positive amplitude ($A_1 = 6.98 \pm 0.66 \times 10^{-8}$) and a phase shift ($\phi_1 = -0.83 \pm 0.10$) while the shorter-period component ($\sim$112 days), characterized by a negative amplitude ($A_2 = -6.72 \pm 0.70 \times 10^{-8}$) and distinct phase shift ($\phi_2 = 0.15 \pm 0.10$). The double-sine model effectively captures both components of the modulation, and the residual analysis shows that this model leaves behind minimal structured noise. Moreover, information criteria (AIC, BIC) and Monte Carlo simulations offer strong statistical support for the reality of the dual QPOs. Notably, the very low $p$-value ($< 0.001$) from the simulation indicates that the observed structure is unlikely to arise by random fluctuations, even in red-noise-dominated light curves \citep{2005A&A...431..391V}.
The  nearly equal modulation strengths of these oscillatory components suggesting the possibility of a beating phenomenon arising from their interference. Such scenario motivates a physical interpretation in terms of a geometric origin such as jet precession or wobbling rather than an intrinsic flux variation. In response, we introduced an amplitude-modulated jet precession model in Section \ref{beat}, wherein the observed flux is modulated by relativistic Doppler boosting due to the periodic variation in jet orientation (jet precession coupled with a secondary polar oscillation). 
This model not only reproduces the double-peaked structure in the frequency domain but also provides a physically coherent explanation for the beat-like amplitude modulation observed in the light curve. In this framework, the double-sine behavior results from the interference of two intrinsic frequencies, \(f_1\) and \(f_2\), associated respectively with jet precession (e.g., induced by Lense–Thirring torques or a supermassive binary black hole system or any other process) and polar modulation. The observed periodicities near 112 and 255 days correspond to the combination frequencies \(f_+ = f_1 + f_2\) and \(f_- = f_1 - f_2\), rather than the jet precession period alone.

The physical picture can be assumed interms of jet precesses as a result of Lense-Thirring torques acting on a tilted accretion disk around a rapidly spinning primary black hole. The precession timescale depends on parameters like black hole spin, inner disk radius, and viscosity. Additionally, the presence of a companion SMBH can further perturb the accretion disk, inducing vertical oscillations or nodding motions (akin to nutation) of the jet axis. These oscillations could be excited by gravitational torques acting on a warped or misaligned inner disk, leading to a second, faster modulation in the jet’s polar angle and hence the Doppler factor.
Together, these two geometric motions - precession and polar oscillation - produce a helical jet structure  with a time-varying pitch and opening angle.  This leads to a modulation in the jet’s viewing angle, and consequently, the observed flux, consistent with the amplitude-modulated beating pattern seen in the light curve. 
PKS 2247–131 provides a compelling example, exhibiting a $\sim$34.5-day $\gamma$-ray QPO that is interpreted as the result of a helical jet geometry, with periodic flux modulation driven by changes in the jet's orientation relative to the observer \citep{2018NatCo...9.4599Z}. In that case, the QPO likely reflects the fundamental of a helical pattern driven by either binary perturbations or intrinsic magnetohydrodynamic (MHD) instabilities. Analogously, in PKS\,0805-07, the observation of the two periodicities ($\sim$255 days and $\sim$112 days) may reflect a superposition of distinct jet-related processes, such as precession, inner-jet oscillations, or higher-order helical harmonics potentially driven by instabilities like kink or Kelvin--Helmholtz modes. This unified scenario explains both periodicities within a physically consistent framework—without the need to invoke an unrealistically compact binary.

Moreover, the windowed LSP analysis revealed that the strength of the detected periodic signals varies over time. This temporal variability indicates that the observed QPOs are transient or episodic rather than persistent, consistent with previous findings in other blazars such as PG~1553+113 \citep{2015ApJ...813L..41A} and OJ~287 \citep{1988ApJ...325..628S,2010MNRAS.402.2087V}. Such intermittency in QPO signatures  further provide indication for evolving jet orientations  \citep[e.g.,][]{2004ApJ...615L...5R,2018MNRAS.474L..81L}.

In conclusion, our multi-technique analysis provides  indication for dual quasi-periodic variability in PKS 0805-07's $\gamma$-ray emission. These results underscore the complexity of blazar variability and suggest that the observed periodicities are likely geometric in origin--possibly due to Lense-Thirring jet precession coupled with a secondary jet-axis oscillation or disk instability. The observed amplitude modulation is consistent with a beating pattern from two near-harmonic signals, supporting a model in which relativistic Doppler boosting modulates the flux due to changes in jet orientation. The QPOs appear episodic rather than persistent—a behavior also observed in other blazars—which may reflect evolving jet dynamics rather than strictly periodic processes.
We emphasize that this interpretation is not unique. The period of approximately 400 days may be attributed to jet precession induced by a SMBBH system. This mechanism has been extensively explored in previous works \citep{2015ApJ...813L..41A, 2017ApJ...836..220C, 2023ApJ...943..157L}. Specifically, jet precession driven by Newtonian dynamics in a close SMBBH system could plausibly account for the observed QPO with a timescale of $\gtrsim$ 1yr. The possibility remains that the two detected QPOs arise from entirely distinct physical processes operating on different spatial or dynamical scales. Future time-resolved, multiwavelength observations and structural studies of the jet will be crucial to disentangle these possibilities. We thus consider our model as one viable scenario among several, and remain open to further refinements or alternative explanations as new data become available.

\acknowledgements
SAD is thankful to the MOMA for the MANF fellowship (No.F.82-27/2019(SA-III)). ZS is supported by the Department of Science and Technology, Govt. of India, under the INSPIRE Faculty grant (DST/INSPIRE/04/2020/002319). SAD, ZS and  NI express  gratitude to the Inter-University Centre for Astronomy and Astrophysics (IUCAA) in Pune, India, for the support and facilities provided.
 
\section{Data Availability}

The data and the model used in this paper will be shared on reasonable request to the contact authors, Sikandar Akbar \& Zahir Shah (e-mail:darprince46@gmail.com or shahzahir4@gmail.com).
\appendix 
\section{Amplitude-Modulated Jet Precession Model}
\label{appen}
The flux $F(t)$ is given by
\begin{equation}\label{eq_a1}
F(t) \propto \delta(t)^p \approx (2\Gamma)^p \left(1 + 2\Gamma^2 p \epsilon \right)
\end{equation}
Substitute $\epsilon$ in \ref{eq_a1}, we get
\begin{equation}\label{eq_a2}
F(t) \propto \delta(t)^p \approx 2^p \Gamma^p \left[1 + 2\Gamma^2 p\left(i\theta \cos\phi - \frac{i^2}{2} - \frac{\theta^2}{2} \right) \right]
\end{equation}\\\\
Using Equation \ref{az_po} in \ref{eq_a2}, and simplifying we get
\begin{align}
F(t) \propto\; & 2^p \Gamma^p \left(1 - \Gamma^2 p \left(i^2 - \frac{\theta_0^2}{2} \right) \right) \Bigg[ \nonumber \\
& 1 + \frac{2\Gamma^2 p i \theta_0}{1 - \Gamma^2 p \left(i^2 - \frac{\theta_0^2}{2} \right)} 
\left( \cos(f_- t + \phi_-) + \cos(f_+ t + \phi_+) \right) \nonumber \\
& - \frac{\Gamma^2 p \theta_0^2}{2 \left(1 - \Gamma^2 p \left(i^2 - \frac{\theta_0^2}{2} \right)\right)} 
\cos(4 f_2 t + 2\phi_2) \Bigg]
\end{align}
\\\\
where,
\begin{equation}
f_\pm = f_1 \pm f_2, \quad
\phi_\pm = \phi_1 \pm \phi_2.
\end{equation}\\\\
Let us define:
\begin{align*}
dF_1 &= \frac{2\Gamma^2 p i \theta_0}{1 - \Gamma^2 p \left(i^2 - \frac{\theta_0^2}{2} \right)}, \quad
dF_2 = \frac{\Gamma^2 p \theta_0^2}{1 - \Gamma^2 p \left(i^2 - \frac{\theta_0^2}{2} \right)} = 2i(dF_1), \\
F_0 &= 2^p \Gamma^p \left(1 - \Gamma^2 p \left(i^2 - \frac{\theta_0^2}{2} \right) \right)
\end{align*}
\\\\
Then, the final flux expression becomes:
\begin{align}
F(t) \propto F_0 \Bigg[1 + \frac{dF_1}{2} \big( \cos((2\pi f_-)t + \phi_-) + \cos((2\pi f_+)t + \phi_+) \big) \nonumber \\
- \frac{dF_2}{2} \cos(4 \pi f_2 t + 2\phi_2) \Bigg]
\end{align}

\bibliographystyle{apsrev4-2} 
\bibliography{apssamp}% Produces the bibliography via BibTeX.

\end{document}